\documentclass[12pt,a4paper,fleqn]{article}
\usepackage[DIV12]{typearea}
\usepackage{amsmath,amsfonts,amssymb}
\usepackage{graphicx}
\usepackage{cite}
\usepackage{mathrsfs}
\usepackage{bbm}
\usepackage{pifont}
\usepackage{units}
\usepackage{longtable}
\usepackage[small,loose]{subfigure}  
\usepackage{pst-all}
\usepackage{cancel}
\usepackage{booktabs}

\usepackage{color}

\newcommand{\Eqref}[1]{equation~\eqref{#1}}

\newcommand{\eVdist}{\kern-0.06em}

\newcommand{\rep}[1]{\ensuremath\boldsymbol{#1}}
\newcommand{\crep}[1]{\ensuremath\overline{\boldsymbol{#1}}}

\DeclareMathOperator{\im}{Im}

\newcommand{\D}{\mathrm{d}}
\newcommand{\I}{\mathrm{i}}

\newcommand{\E}[1]{\ensuremath{\mathrm{E}_{#1}}} 

\newcommand{\SO}[1]{\ensuremath{\mathrm{SO}(#1)}}
\newcommand{\SU}[1]{\ensuremath{\mathrm{SU}(#1)}}
\newcommand{\U}[1]{\ensuremath{\mathrm{U}(#1)}}
\newcommand{\Z}[1]{\ensuremath{\mathbbm{Z}_{#1}}} 

\numberwithin{equation}{section}
\numberwithin{table}{section}

\allowdisplaybreaks[1]

\title{String-derived MSSM vacua with residual $\boldsymbol{R}$ symmetries}

\begin{document}

\begin{titlepage}
\vspace*{-1cm}
\begin{flushright}
TUM-HEP 787/10\\
MPP-2010-169\\
NSF-KITP-10-160\\
LMU-ASC 106/10 \\
OHSTPY-HEP-T-10-006
\end{flushright}

\vspace*{0.5cm}

\begin{center}
{\Large\bf
String-derived MSSM vacua with residual $\boldsymbol{R}$ symmetries
}

\vspace{0.7cm}

\textbf{
Rolf Kappl
$^{\,a,b}${},
Bj\"orn Petersen
$^{\,a}${},
Stuart Raby
$^{\,c,d}${},
Michael Ratz
$^{\,a,d}${},
Roland Schieren
$^{\,a}${},
Patrick~K. S. Vaudrevange
$^{\,e}${}
}
\\[8mm]
\textit{$^a$\small
~Physik-Department T30, Technische Universit\"at M\"unchen, \\
James-Franck-Stra\ss e, 85748 Garching, Germany
}
\\[5mm]
\textit{$^b$\small
~Max-Planck-Institut f\"ur Physik (Werner-Heisenberg-Institut),\\
F\"ohringer Ring 6,
80805 M\"unchen, Germany
}
\\[5mm]
\textit{$^c$\small
~Department of Physics, The Ohio State University,\\
191 W.\ Woodruff Ave., Columbus, OH 43210, USA
}
\\[5mm]
\textit{$^d$\small
~Kavli Institute for Theoretical Physics,
University of California,\\
Santa Barbara, CA 93106-4030
}
\\[5mm]
\textit{$^e$\small
~Arnold Sommerfeld Center for Theoretical Physics,\\
Ludwig-Maximilians-Universit\"at M\"unchen, 80333 M\"unchen, Germany
}
\end{center}

\vspace{0.4cm}

\begin{abstract}
Recently it was shown that there is a unique $\Z4^R$ symmetry for the MSSM which
allows the Yukawa couplings and dimension five neutrino mass operator, forbids
the $\mu$ term and commutes with SO(10). This $\Z4^R$ symmetry contains matter
parity as a subgroup and forbids dimension four and five proton decay operators.
We show how to construct string vacua with discrete $R$ symmetries in general
and this symmetry in particular, and present an explicit example which exhibits
the exact MSSM spectrum, the $\Z4^R$ symmetry as well as  other desired features
such as gauge-top unification. We introduce the Hilbert basis method for
determining all $D$-flat configurations and efficient algorithms for identifying
field configurations with a desired residual symmetry.  These methods are used
in an explicit example, in which we  describe in detail how to construct a
supersymmetric vacuum configuration with the phenomenologically attractive
$\Z4^R$ symmetry. At the perturbative level, this is a supersymmetric Minkowski
vacuum in which almost all singlet fields (moduli) are fixed.
\end{abstract}

\end{titlepage}

\newpage

\section{Introduction}

There are many independent observations hinting at the relevance of a high scale
for particle physics. The smallness of neutrino masses has a simple explanation
in terms of the see-saw mechanism~\cite{Minkowski:1977sc} and relies on the
existence of heavy singlet neutrinos. Stabilizing the electroweak scale against
the see-saw scale seems to require supersymmetry; remarkably, the simplest
supersymmetric extension of the standard model, the MSSM, realizes the
compelling scenario of gauge unification~\cite{Dimopoulos:1981yj} at a scale
$M_\mathrm{GUT}\simeq2\cdot10^{16}\,\mathrm{GeV}$, which is suspiciously close
to the see-saw scale. Both scales are not too far from $M_\mathrm{P}\simeq2\cdot
10^{18}\,\mathrm{GeV}$, which is set by Newton's constant.

The question of how to incorporate all scales in a coherent scheme has been
addressed for more than 30 years. From a bottom-up perspective one is led to the
scheme of grand unified theories (GUTs). Although this scheme exhibits various
very appealing features, there are three major obstacles. First, there is the
so-called doublet-triplet splitting, and related to it, the MSSM $\mu$ problem.
Second, even if this problem is solved, unified models typically are in conflict
with dimension five proton decay \cite{Sakai:1981pk,Dimopoulos:1981dw}
operators.  (It is well known that dimension four proton decay can be forbidden
by matter parity.\footnote{Matter parity is sometimes also known as ``$R$
parity''. We  choose not to use this terminology as matter parity is non-$R$,
i.e.\ it commutes with supersymmetry, and one point of this paper is to discuss
a true discrete $R$ symmetry.}) Third, in four-dimensional models of grand
unification there is no relation between the GUT and Planck scales,
$M_\mathrm{GUT}$ and $M_\mathrm{P}$.

String theory is believed to provide us with such a relation. However, if string
theory is to describe the real world, it should also provide us with solutions
to the first and second problems. In fact, as known for a long time, the
doublet-triplet splitting problem has a simple solution in theories with extra
dimensions in which the GUT symmetry is broken in the process of
compactification \cite{Witten:1985xc,Breit:1985ud}, which also avoids the most
stringent problems with dimension five proton decay \cite{Altarelli:2001qj}.
However, in concrete string compactifications  (see
\cite{Ibanez:1987sn,Font:1988mm,Font:1989aj} for early attempts and
\cite{Cleaver:1998saa,Cleaver:1999mw} for a different approach) very often the problem is
reintroduced; this applies also to the models discussed more recently
\cite{Buchmuller:2005jr,Buchmuller:2006ik,Lebedev:2006kn,Lebedev:2007hv}. On the
other hand, one cannot rule out these constructions as their vacua are not
completely understood. That is, the analysis of potentially realistic string
models is a non-trivial task since a given model exhibits a plethora of vacua
with very different features. The role of discrete symmetries in identifying and
analyzing such vacua has been stressed recently \cite{Buchmuller:2008uq}. One of
these symmetries is matter parity, which has been successfully embedded in string
theory~\cite{Lebedev:2007hv}. This study is devoted to a discussion of the role
of further discrete symmetries in such models and their phenomenological implications.
Specifically, we will focus on a proposed $\Z4^R$ symmetry
\cite{Babu:2002tx}
which has recently been shown in~\cite{Lee:2010gv} to be the unique anomaly-free possibility
with the following properties:
\begin{enumerate}
 \item it forbids the $\mu$ term at the perturbative level;
 \item it allows the MSSM Yukawa couplings and the effective neutrino mass operator;
 \item it commutes with SO(10) in the matter sector.
\end{enumerate}
This symmmetry has the appealing feature that it forbids automatically dimension
four and five proton decay operators.  We will discuss how to identify string
vacua exhibiting this symmetry and present a globally consistent string-derived
model with the exact MSSM spectrum realizing this symmetry.

We start in section~\ref{sec:GeneralPicture} with a short description of the
general picture. In section~\ref{sec:the_model} we present an explicit
string-derived model in which an anomalous $\Z4^R$
symmetry explains a suppressed vacuum expectation value of the superpotential,
provides us with a solution of the $\mu$ problem and suppresses dimension five
proton decay operators. Section~\ref{sec:Summary} contains our conclusions. In
various appendices we collect details of our calculations.

\section{General picture}
\label{sec:GeneralPicture}

String theory compactifications provide us with a plethora of vacuum
configurations, each of which comes with symmetries and, as a consequence, with
extra massless degrees of freedom whose mass terms are prohibited by these symmetries.
Simple examples for such compactifications include heterotic
orbifolds~\cite{Dixon:1985jw,Dixon:1986jc}, where the rank of the gauge group
after compactification equals that of $\E8\times\E8$, i.e.\ 16. A few hundreds
of orbifold models are known in which $\E8\times\E8$ gets broken to the standard
model gauge symmetry
$G_\mathrm{SM}=\SU3_C\times\SU2_\mathrm{L}\times\U1_Y$ (with
hypercharge in GUT normalization) times $\U1^n$ times a hidden sector group  and
the chiral spectra of the MSSM \cite{Lebedev:2006kn,Lebedev:2008un}. They also
exhibit exotics which are vector-like with respect to $G_\mathrm{SM}$ and which can be
decoupled when the extra gauge symmetries are broken. Each of these models
contains many vacua, i.e.\ solutions of the supersymmetry conditions
$V_F=V_D=0$. Typically these vacua exhibit flat directions before supersymmetry
breaking.

At the orbifold point, where the vacuum expectation values (VEVs) of all fields
are zero,  we have discrete $R$ as well as continuous and discrete non-$R$
symmetries. Typically one of the \U1 symmetries appears anomalous, which is
conventionally denoted by $\U1_\mathrm{anom}$. Also some of the discrete
symmetries may appear anomalous \cite{Ibanez:1992hc,Araki:2008ek}. After
assigning  VEVs to certain fields, some of the symmetries are spontaneously
broken and others remain. We shall be mainly interested in remnant discrete
symmetries, which can be of $R$ or non-$R$ type and be either anomalous or
non-anomalous. We will discuss examples of all kinds in
section~\ref{sec:the_model}.

Clearly, one cannot assign VEVs to the fields at will. Rather, one has to
identify field configurations which correspond to local minima of the
(effective) scalar potential. Let us briefly describe the first steps towards
identifying such vacuum configurations. Consider a configuration in which
several fields attain VEVs. We focus on ``maximal vacua'' (as
in~\cite{Buchmuller:2008uq}), i.e.\ we assume that all fields which are neutral
under the remnant gauge and discrete symmetries, called $\phi^{(i)}$ $(1\le i\le
N)$ in what follows, attain VEVs (if these are consistent with
$D$-flatness).  All fields without expectation value, denoted
by $\psi^{(j)}$ $(1\le j\le M)$, therefore transform non-trivially under some of
the remnant symmetries.

\subsection{Discrete non-$\boldsymbol{R}$ symmetries}

The case of vacua with non-$R$ discrete symmetries has
been discussed in detail in \cite{Buchmuller:2006ik,Buchmuller:2008uq}.
In this case, the superpotential has the form
\begin{equation}
 \mathscr{W}~=~\Omega(\phi^{(1)},\dots\phi^{(N)})+(\text{terms at least
 quadratic in the}~\psi^{(j)})\;.
\end{equation}
Therefore, the $F$-term equations for the $\psi^{(j)}$ fields trivially vanish and we are
left with $N$ $F$-term equations for the $N$ $\phi^{(i)}$ fields, which generically
have solutions. Hence, if all $\phi^{(i)}$ enter gauge invariant monomials
composed of $\phi^{(i)}$ fields only, we will find supersymmetric vacua, i.e.\
solutions to the $F$- and $D$-term equations.

Because of the above arguments it is sufficient to look at the system of
$\phi^{(i)}$ fields only, which has been studied in the literature. Consider the
case of a \emph{generic} superpotential $\mathscr{W}$. It is known that the
solutions to the $D$- and $F$-term equations intersect \emph{generically} in a
point~\cite{Luty:1995sd}. That is, there are point-like field configurations
which satisfy
\begin{equation}\label{eq:GenericVacuum}
 D_a~=~F_i~=~0\quad\text{at}~\phi^{(i)}~=~\langle\phi^{(i)}\rangle\;,
\end{equation}
where, as usual,
\begin{subequations}
\begin{eqnarray}
 D_a & = & \sum_i (\phi^{(i)})^*\,\mathsf{T}_a\,\phi^{(i)}\;,\\
 F^{(i)} & = & \frac{\partial\mathscr{W}}{\partial\phi^{(i)}}\;.
\end{eqnarray}
\end{subequations}
The term `point-like' means that there are no massless deformations of the
vacuum \eqref{eq:GenericVacuum}. The reason why these vacua are point-like is
easily understood: generically the $F$-term equations constitute as many gauge
invariant constraints as there are gauge invariant variables. However, this also
means that, at least generically,
\begin{equation}
 \mathscr{W}|_{\phi^{(i)}~=~\langle\phi^{(i)}\rangle}~\ne~0\;.
\end{equation}
If the fields attain VEVs $\langle\phi^{(i)}\rangle$ of the order of the
fundamental scale, one hence expects to have too large a VEV for $\mathscr{W}$.
One possible solution to the problem relies on approximate $R$
symmetries~\cite{Kappl:2008ie}, where one obtains a highly suppressed VEV of the
superpotential. In what follows, we discuss an alternative: in settings with a
residual $R$ symmetry the above conclusion can be avoided as well.

\subsection{Discrete $\boldsymbol{R}$ symmetries}

Let us now discuss vacua with discrete $R$ symmetries. To be
specific, consider the order four symmetry $\Z4^R$, under which the
superpotential $\mathscr{W}$ has charge 2, such that
\begin{equation}
 \mathscr{W}~\xrightarrow{\zeta}~-\mathscr{W}
\end{equation}
under the $\Z4^R$ generator $\zeta$. Superspace coordinates transform as
\begin{equation}
 \theta_\alpha~\to~\I\,\theta_\alpha
\end{equation}
such that the $F$-term Lagrangean
\begin{equation}
 \mathscr{L}_F~=~\int\!\D^2\theta\,\mathscr{W}+\text{h.c.}
\end{equation}
is invariant. Chiral superfields will have $R$ charges $0,1,2,3$.\footnote{A
special role is played by the dilaton $S$, whose imaginary part $a=\im
S|_{\theta=0}$ shifts under $\Z4^R$.} Both the fields of the type $\psi_1$ and
$\psi_3$ with $R$ charges $1$ and $3$, respectively, can
acquire mass as the $\psi_1^2$ and $\psi_3^2$ terms have $R$ charge $2\mod4$ and
thus denote allowed superpotential terms.

The system of fields $\phi_0^{(i)}$ and $\psi_2^{(j)}$ with $R$ charges 0 and 2,
respectively, is more interesting. Consider first only one field $\phi_0$ and
one field $\psi_2$. The structure of the superpotential is
\begin{equation}\label{eq:superpotential0+2}
 \mathscr{W}~=~\psi_2\cdot f(\phi_0)+\mathcal{O}(\psi_2^3)
\end{equation}
with some function $f$. The $F$-term for $\phi_0$ vanishes trivially as long as
$\Z4^R$ is unbroken,
\begin{equation}
 \frac{\partial\mathscr{W}}{\partial\phi_0}
 ~=~
 \psi_2\cdot f'(\phi_0)
 ~=~0\;.
\end{equation}
Note that due to the $\Z4^R$ symmetry the superpotential vanishes in the
vacuum. Thus it is sufficient to look at the global supersymmetry $F$-terms. On the
other hand, the $F$-term constraint (at $\psi_2=0$)
\begin{equation}
 \frac{\partial\mathscr{W}}{\partial\psi_2}
 ~=~f(\phi_0)~\stackrel{!}{=}~0
\end{equation}
will in general fix $\phi_0$ at some non-trivial zero $\langle\phi_0\rangle$ of
$f$. Indeed, there will be a supersymmetric mass term, which can be seen by
expanding $\phi_0$ around its VEV, i.e.\ inserting
$\phi_0=\langle\phi_0\rangle+\delta\phi_0$ into
\eqref{eq:superpotential0+2},
\begin{equation}
 \mathscr{W}~=~f'(\langle\phi_0\rangle)\,\delta\phi_0\,\psi_2
 +\mathcal{O}(\delta\phi_0^2,\psi_2^3)\;.
\end{equation}
The supersymmetric mass $f'(\langle\phi_0\rangle)$ is generically
different from 0.

Repeating this analysis for $N$ $\phi_0^{(i)}$ and $M$ $\psi_2^{(j)}$ fields
reveals that the $F$-terms of the $\psi_2^{(j)}$ lead to $M$, in general
independent, constraints on the $\phi_0^{(i)}$ VEVs. For $N=M$ we therefore
expect point-like vacua with all directions fixed in a supersymmetric way.

To summarize, systems with a residual $R$ symmetry ensure, unlike in the case
without residual symmetries, that $\langle\mathscr{W}\rangle=0$. However, in
systems which exhibit a linearly realized $\Z4^R$ somewhere in field space it
may not be possible to find a supersymmetric vacuum that preserves $\Z4^R$. In
the case of a generic superpotential this happens if there are more, i.e.\
$M>N$, fields with $R$ charge 2 than with 0. On the other hand, if there are
more fields with $R$ charge 0 than with 2, i.e.\ for $M<N$, one expects to have
a Minkowski vacuum with $N-M$ flat directions. For $N=M$ one can have
supersymmetric Minkowski vacua with all directions fixed in a supersymmetric
way.

An important comment in this context concerns the moduli-dependence of
couplings. As we have seen, in the case of discrete $R$-symmetries one might
obtain more constraint (i.e.\ $F$-term) equations than $R$-even `matter' fields.
Specifically, in string vacua one should, however, carefully take into account
all $R$-even fields, also the K\"ahler and complex structure moduli, $T_i$ and
$U_j$, on whose values the coupling strengths depend.

\section{An explicit string-derived model}
\label{sec:the_model}

In order to render our discussion more specific, we base our analysis on a
concrete model. We consider a $\Z2\times\Z2$ orbifold compactification with an
additional freely acting  $\Z2$ of the $\E8\times\E8$ heterotic string. Details
of the model including shift vectors and Wilson lines can be found in
appendix~\ref{app:details_model}.

In \cite{Blaszczyk:2009in} a vacuum configuration of a very similar
$\Z2\times\Z2$ model with matter parity and other desirable features was
presented. However, the vacuum configuration discussed there has the unpleasant
property that, at least generically, all Higgs fields attain large masses. In
what follows we discuss how this can be avoided by identifying vacuum
configurations with enhanced symmetries. In \cite{Lee:2010gv} another vacuum
with the $\Z4^R$ symmetry discussed in the introduction was found by using the
methods presented in this paper. In both models the GUT symmetry
is broken non-locally. This may be advantageous from the point of view of
precision gauge unification \cite{Hebecker:2004ce}. It also avoids fractionally
charged exotics, which appear in many other compactifications (cf.\ the
discussion in \cite{GatoRivera:2010xn}).

\paragraph{Labeling of states.}
We start our discussion with a comment on our notation. In a first step, we
label the fields according to their
$G_\mathrm{SM}\times[\SU3\times\SU2\times\SU2]_\mathrm{hid}$ quantum numbers. In
particular, we denote the standard model representations with lepton/Higgs and
$d$-quark quantum numbers as
\begin{subequations}
\begin{eqnarray}
 L_i & : & (\boldsymbol{1},\boldsymbol{2})_{-1/2}\;,\\
 \bar L_i & : & (\boldsymbol{1},\boldsymbol{2})_{1/2}\;,\\
 D_i & : & (\boldsymbol{3},\boldsymbol{1})_{-1/3}\;,\\
  \bar D_i & : & (\boldsymbol{\overline{3}},\boldsymbol{1})_{1/3}\;.
\end{eqnarray}
\end{subequations}
In the next step we identify $\Z4^R$ such that the $\bar L_i$/$L_i$ decompose in
lepton doublets $\ell_i$ with odd $\Z4^R$ charges and Higgs candidates 
$h_d$/$h_u$ with even $\Z4^R$ charges etc. The details of labeling states are
given in appendix~\ref{app:details_model}.

\paragraph{Searching for $\boldsymbol{\Z4^R}$.}
How can one obtain vacua with $\Z4^R$ in practice? We found
the following strategy most efficient:
\begin{enumerate}
\item In a first step we switch on a random sample of SM singlets in such a way
that all unwanted gauge factors are spontaneously broken.
\item With these VEVs at hand, the original gauge and discrete symmetries at the
orbifold point get broken to a discrete subgroup, which can be determined
unambiguously with the methods described in \cite{Petersen:2009ip}. Details of
the automatization of these methods are explained in \cite{Schieren:2010pt}.
\item We only keep configurations in which there is a residual $\Z4^R$ symmetry
with precisely three generations of matter having $R$-charge 1; details of how
to identify such configurations are given in appendix~\ref{app:IdentifyingZ4R}.
\item From these configurations we select those exhibiting the following properties:
\begin{itemize}
 \item $F$- and $D$-flat;
 \item all exotics decouple;
 \item one pair of massless Higgs, i.e.\ $\mu$ term forbidden to all orders (at
 the perturbative level);
 \item Yukawa couplings have full rank.
\end{itemize}
\end{enumerate}
One of the main achievements of this study is a considerable simplification in
the verification of the four items listed in step 4. In order to check
$D$-flatness of a given configuration we use the Hilbert basis method, which is
described in detail in appendix~\ref{app:HilbertBasis}. The other three
properties can be verified by inspecting the remnant discrete symmetries only.
In earlier studies
\cite{Buchmuller:2005jr,Buchmuller:2006ik,Lebedev:2006kn,Lebedev:2007hv} we had
to explicitly identify couplings that are consistent with the string selection
rules in order to show that all exotics decouple and the Yukawa couplings have
full rank. In our new approach the remnant symmetries will tell us immediately
whether an entry of a mass or Yukawa matrix will or will not appear. We have
cross-checked this method extensively by explicitly computing the couplings
between the charged and the VEV fields, and were always able to find a coupling
which fills in an entry of a matrix, albeit sometimes at very high orders. Note,
we assume that all couplings allowed by string selection rules appear in the superpotential.

\paragraph{VEV configuration.}
Following the above steps, we obtained a promising
configuration in which the fields
\begin{eqnarray}
 \widetilde{\phi}^{(i)}
 & = &
 \{\phi _1, \phi _2, \phi _3, \phi _4, \phi _5, \phi _6, \phi _7, \phi _8,
 \phi _9, \phi _{10}, \phi _{11}, \phi _{12}, \phi _{13}, \phi _{14}, \nonumber\\
 &   & \hphantom{\{}{}
 x_1,x_2,x_3,x_4,x_5,\bar x_1,\bar x_3 ,\bar x_4 ,\bar x_5,y_3,y_4,y_5,y_6\}
\label{eq:phitildefields}
\end{eqnarray}
attain VEVs. The full quantum numbers of these fields are given
in table~\ref{tab:fullspectrum} in appendix~\ref{app:details_model}. In order
to ensure $D$-flatness with respect to the hidden sector gauge factors, in a given basis
not all components of the $x_i$/$\bar x_i$ and $y_i$ attain VEVs. Details are
given in equations \eqref{eq:SU2vacuum} and \eqref{eq:SU3vacuum} in
appendix~\ref{app:SUSYVacuum}.

\paragraph{Remnant discrete symmetries.} By giving VEVs to the
$\widetilde{\phi}^{(i)}$ fields in \eqref{eq:phitildefields}, we arrive  at a
vacuum in which, apart from $G_\mathrm{SM}$ and a `hidden' \SU2, all gauge
factors are spontaneously broken. The vacuum exhibits a
$\Z4^R$ symmetry, whereby the superpotential $\mathscr{W}$
has $\Z4^R$ charge 2.

The $\Z4^R$ charges of the matter fields are shown in
table~\ref{tab:DiscreteChargesMatterZ2xZ2_1-1}. The detailed origin of the $\Z4^R$
symmetry is discussed later. Given these charges, we confirm
by a straightforward field-theoretic calculation
(cf.~\cite{Ibanez:1991hv,Araki:2008ek}) that $\Z4^R$ appears indeed anomalous with
universal  $\SU2_\mathrm{L}-\SU2_\mathrm{L}-\Z4^R$ and  $\SU3_C-\SU3_C-\Z4^R$
anomalies (see \cite{Lee:2010gv} and
appendix~\ref{app:DiscreteAnomalyCalculation}). The statement
that $\Z4^R$ appears anomalous
means, as we shall discuss in detail below, that the
anomalies are cancelled by a Green-Schwarz (GS) mechanism.
On the other hand, the $\Z4^R$ has a, by the traditional
criteria, non-anomalous
$\Z2^\mathcal{M}$ subgroup which is equivalent to matter
parity~\cite{Lee:2010gv}.

\begin{table}[htb]
\centering
\subtable[Quarks and leptons.]{
\begin{tabular}{cccccc}
\toprule[1.3pt]
 & $q_i$ & $\bar u_i$ & $\bar d_i$ & $\ell_i$ & $\bar e_i$ \\
$\Z4^R$ & 1 & 1 & 1 & 1 & 1\\
\bottomrule[1.3pt]
\end{tabular}
}
\\
\subtable[Higgs and exotics.]{
\begin{tabular}{ccccccccccccccccccc}
\toprule[1.3pt]
 & $h_1$ & $h_2$& $h_3$& $h_4$& $h_5$& $h_6$ & $\bar h_1$ & $\bar h_2$ & $\bar h_3$ & $\bar h_4$ & $\bar h_5$ &  $\bar h_6$ & $\delta_1$ & $\delta_2$ & $\delta_3$ & $\bar\delta_1$ & $\bar\delta_2$ & $\bar\delta_3$ \\
$\Z4^R$ & 0 & 2 & 0 & 2 & 0 & 0 & 0 & 2 & 0 & 0 & 2 & 2 & 0 & 2 & 2 & 2 & 0 & 0\\
\bottomrule[1.3pt]
\end{tabular}
}
\caption{$\Z4^R$ charges of the (a) matter fields and (b) Higgs and exotics. The
index $i$ in (a) takes values $i=1,2,3$.}
\label{tab:DiscreteChargesMatterZ2xZ2_1-1}
\end{table}

\paragraph{$\boldsymbol{D}$-flatness.}
As already discussed, we cannot switch on the $\widetilde{\phi}^{(i)}$ fields at will; rather we have to show that there are
vacuum configurations in which all these fields acquire VEVs. This requires to
verify that the $D$- and $F$-term potentials vanish.
With the Hilbert basis method (see appendix~\ref{app:HilbertBasis}) we could
identify a complete set of $D$-flat directions composed of $\widetilde{\phi}^{(i)}$
fields. We compute the dimension of the $D$-flat moduli space
using Singular \cite{GPS05} and the STRINGVACUA \cite{Gray:2008zs} package; the
result is that there are $18$ $D$-flat directions; the details of the
computation are collected in appendix~\ref{app:SUSYVacuum}.

\paragraph{$\boldsymbol{F}$-term constraints.} Next we consider the $F$-term
constraints. As discussed in section~\ref{sec:GeneralPicture}, the $F$-term
conditions come from the fields with $R$-charge 2. We compute the number of
independent conditions in appendix~\ref{app:SUSYVacuum}. The result is that
there are 23 independent conditions on $18+6=24$ $D$-flat directions, where we
included the K\"ahler and complex structure moduli. We therefore expect to find
supersymmetric vacuum configurations in which all the $\widetilde{\phi}^{(i)}$
acquire VEVs. In this configuration, almost all singlet fields, including the
geometric moduli are fixed in a supersymmetric way. It will be interesting to
compare this result to similar results found recently in the context of smooth
heterotic compactifications \cite{Anderson:2010mh}. We expect a significantly
different, i.e.\ healthier, phenomenology than in the case in which a large
number of singlets acquire mass only after supersymmetry breaking~\cite{Dundee:2010sb,Parameswaran:2010ec}.
Notice that there are two possible caveats. First, the analysis
performed strictly applies only to superpotentials which are, apart from all
the symmetries we discuss, generic. Second, it might happen that there are
supersymmetric vacua, but they occur at large VEVs of some of the
fields, i.e.\ in regions of field space where we no longer control our
construction. Both issues will be addressed elsewhere.

\paragraph{Higgs vs.\ matter.}
The $\Z2^\mathcal{M}$ subgroup of the $\Z4^R$ symmetry allows us to discriminate between
\begin{itemize}
\item  3 lepton doublets, $\ell_i=\{L_4, L_6, L_7\}$,
\item  3 $d$-type quarks,
 $\bar d_i=\{\bar D_1, \bar D_3, \bar D_4\}$,
\end{itemize}
on the one hand, and
\begin{itemize}
\item Higgs candidates, $h_i=\{L_1,L_2,L_3,L_5,L_8,L_9\}$ and
 $\bar h_i=\{\bar L_1,\bar L_2,\bar L_3,\bar L_4,\bar L_5,\bar L_6\}$,
\item exotic triplets, $\delta_i=\{D_1,D_2,D_3\}$ and
 $\bar\delta_i=\{\bar D_2,\bar D_5,\bar D_6\}$
\end{itemize}
on the other hand.

\paragraph{Decoupling of exotics.}
With the charges in table~\ref{tab:DiscreteChargesMatterZ2xZ2_1-1} we can
readily analyze the structure of the mass matrices.
We crosscheck these structures by explicitly computing the
couplings allowed by the string selection rules (cf.\ \cite{Blaszczyk:2009in}). Note there is a caveat: our results are based on the
assumption that all couplings that are allowed by the selection
rules will appear with a non-vanishing coefficient.
A $\widetilde{\phi}^{n}$ in the matrices
represents a known polynomial of order $n$ in the $\widetilde{\phi}$ fields which we
have calculated using string selection rules. A zero entry in the matrices means
that the corresponding coupling is not present in the perturbative
superpotential. The $\bar h_i-h_j$ Higgs mass matrix is
\begin{equation}
 \mathcal{M}_h  ~=~
\begin{pmatrix}
 0 & \phi _6 & 0 & \phi _4 & 0 & 0 \\
 \phi _7 & 0 & \phi _2 & 0 & \phi _{13} & \phi _{14} \\
 0 & \phi _1 & 0 & \widetilde{\phi}^3 & 0 & 0 \\
 0 & \widetilde{\phi}^3& 0 & \widetilde{\phi}^5 & 0 & 0 \\
 \widetilde{\phi}^3 & 0 & \phi _{11} & 0 & \phi _8 &\widetilde{\phi}^3 \\
 \widetilde{\phi}^3 & 0 & \phi _{12} & 0 & \widetilde{\phi}^3 & \phi _8
\end{pmatrix}\;.
\end{equation}
Here we omit coefficients, which depend on the three K\"ahler
moduli $T_i$ and complex structure moduli $U_i$.
Clearly, this mass matrix has rank five, such that there is one massless Higgs
pair
\begin{subequations}
\begin{eqnarray}
 h_u & = & a_1\, \bar h_1 +a_2\, \bar h_3 + a_3 \bar h_4\;,\\
 h_d & = & b_1\,h_1 +b_2\,h_3 +b_3\,h_5 +b_4\,h_6
\end{eqnarray}
\end{subequations}
with $a_i$ and $b_j$ denoting coefficients.
The $\bar \delta-\delta$ mass matrix is
\begin{equation}
 \mathcal{M}_\delta ~=~
\left(
\begin{array}{ccc}
\widetilde{\phi}^5 & 0 & 0 \\
 0 & \phi _8 & \widetilde{\phi}^3\\
 0 & \widetilde{\phi}^3 & \phi _8
\end{array}
\right)\;.
\end{equation}
Hence, the matrix has full rank and all exotics decouple. Note that the block
structure of $\mathcal{M}_\delta$ is not a coincidence but a consequence of the
fact that $\delta_2/\delta_3$ and $\bar \delta_2/\bar \delta_3$ form  $D_4$
doublets (see below). Altogether we see that all exotics with
Higgs quantum numbers, and all but one pair of exotic triplets, decouple at the
linear level in the $\widetilde{\phi}^{(i)}$ fields. This leads to the
expectation that all but one pair of exotics get mass of the order of the GUT
(or compactification) scale $M_\mathrm{GUT}$ while one pair of triplets might be
somewhat lighter. We also note that the presence of colored states somewhat
below $M_\mathrm{GUT}$ can give a better fit to MSSM gauge coupling unification
(cf.\ \cite{Dundee:2008ts}). However, a crucial property of the $\delta$- and
$\bar\delta$ triplets is that, due to the $\Z4^R$ symmetry, they do not mediate
dimension five proton decay.

\paragraph{Effective Yukawa couplings.}
The effective Yukawa couplings are defined by
\begin{equation}
 \mathscr{W}_Y~=~
 \sum_{i=1,3,4}\left[ (Y_u^{(i)})^{fg}\,q_f\,\bar u_g\,\bar h_i \right]
 +\sum_{i=1,3,5,6}
 \left[(Y_d^{(i)})^{fg}\,q_f\,\bar d_g\,h_i
 +(Y_e^{(i)})^{fg}\,\ell_f\,\bar e_g\,h_i\right]\;.
\label{eq:definition_yukawas}
\end{equation}
The Yukawa coupling structures are
\begin{subequations}\label{eq:YukawaMatricesZ2xZ2_1-1}
\begin{eqnarray}
Y_u^{(1)} & = &
\left(
\begin{array}{ccc}
 \widetilde{\phi} ^2 &  \widetilde{\phi} ^4 & \widetilde{\phi} ^6 \\
  \widetilde{\phi} ^4 & \widetilde{\phi} ^2 & \widetilde{\phi} ^6 \\
 \widetilde{\phi} ^6 & \widetilde{\phi} ^6 & 1
\end{array}
\right) \;,
\quad
Y_u^{(3)}  ~=~
\left(
\begin{array}{ccc}
 1 &  \widetilde{\phi} ^6 & \widetilde{\phi} ^4 \\
  \widetilde{\phi} ^6 & 1 & \widetilde{\phi} ^4 \\
 \widetilde{\phi} ^4 & \widetilde{\phi} ^4 & \widetilde{\phi} ^2
\end{array}
\right)\;,
\label{eq:Yu}\\
Y_e^{(5)} =  (Y_d^{(5)})^T & = &
\left(
\begin{array}{ccc}
 \widetilde{\phi} ^6 & \widetilde{\phi} ^6 &  \widetilde{\phi} ^6 \\
 \widetilde{\phi} ^6 & \widetilde{\phi} ^6 & 1 \\
  \widetilde{\phi} ^6 & 1 & \widetilde{\phi} ^4
\end{array}
\right) \;,
\\
Y_e^{(6)} =  (Y_d^{(6)})^T & = &
\left(
\begin{array}{ccc}
 \widetilde{\phi} ^6 & \widetilde{\phi} ^6 & 1 \\
 \widetilde{\phi} ^6 & \widetilde{\phi} ^6 &  \widetilde{\phi} ^6 \\
 1 &  \widetilde{\phi} ^6 & \widetilde{\phi} ^4
\end{array}
\right) \;.
\label{eq:YdYe}
\end{eqnarray}
\end{subequations}
$Y_d$ and $Y_e$ coincide at tree-level, i.e.\ they exhibit \SU5
GUT relations, originating from the non-local GUT breaking due to the freely acting Wilson line.
There are additional contributions to $Y_u$ from couplings to $\bar h_4$ and to $Y_e$/$Y_d$
from couplings to $h_{1,3}$ which can be neglected if the VEVs of the
$\widetilde{\phi}^{(i)}$ fields are small.

Because of the localization of the matter fields, we expect the renormalizable
(1,3) and (3,1) entries in $Y_e^{(6)}$ to be exponentially suppressed.

\paragraph{Gauge-top unification.} The $(3,3)$ entry of $Y_u$ is related to the
gauge coupling. More precisely, in an orbifold GUT limit in which the first
$\Z2$ orbifold plane is larger than the other dimensions there is an \SU6 bulk
gauge symmetry, and the ingredients of the top Yukawa coupling $h_u$ (i.e.\ the
fields $\bar h_{1,3,4}$), $\bar u_3$ and $q_3$ are bulk fields of this plane,
i.e.\ hypermultiplets in the $N=2$ supersymmetric description. As discussed in
\cite{Hosteins:2009xk}, this implies that the top Yukawa coupling $y_t$ and the
unified gauge coupling $g$ coincide at tree-level. Moreover, localization
effects in the two larger dimensions \cite{Lee:2003mc} will lead to a slight
reduction of the prediction of $y_t$ at the high scale such that realistic top
masses can be obtained.

\paragraph{$\boldsymbol{D_4}$ flavor symmetry.} The block structure of the
Yukawa matrices is not a coincidence but a consequence of a $D_4$ flavor
symmetry \cite{Ko:2007dz},  related to the vanishing Wilson line in the $e_1$
direction, $W_1 = 0$ (cf.\ e.g.\ \cite{Kobayashi:2006wq}).  The first two
generations transform as a $D_4$ doublet, while the third generation is a $D_4$
singlet.

\paragraph{Neutrino masses.} In our model we have 11 neutrinos,
i.e.\ SM singlets whose charges are odd under $\Z4^R$ meaning that
they have odd $\Z2^\mathcal{M}$ charge, where
$\Z2^\mathcal{M}$ is the matter parity subgroup of $\Z4^R$. Their
mass matrix has rank 11 at the perturbative level. The neutrino
Yukawa coupling is a $3\times 11$ matrix and has full rank. Hence
the neutrino see-saw mechanism with many neutrinos
\cite{Buchmuller:2007zd} is at work.

\paragraph{Proton decay operators.} The $\Z4^R$ symmetry forbids
all dimension four and five proton decay operators at the
perturbative level~\cite{Lee:2010gv}. In addition, the
non-anomalous matter parity subgroup $\Z2^\mathcal{M}$ forbids all
dimension four operators also non-perturbatively. The dimension
five operators like $q\,q\,q\,\ell$ are generated
non-pertubatively, as we will discuss below.

\paragraph{Non-perturbative violation of $\boldsymbol{\Z4^R}$.}
Once we include the terms that are only
forbidden by the $\Z4^R$ symmetry, we obtain further couplings.
An example for such an additional term is the dimension five proton decay operator,
\begin{equation}\label{eq:Wnp1}
 \mathscr{W}_{np}
 ~\supset~
 q_1\,q_1\,q_2\,\ell_1\,
 \mathrm{e}^{-a\,S}\, (x_4 \bar x_5 + x_5 \bar x_4)
\left[\begin{pmatrix}\phi_{11}\\\phi_{12}\end{pmatrix} \cdot \begin{pmatrix}\phi_{11}\\\phi_{12}\end{pmatrix}\right]^3
\phi_4\,\phi_7^2
\left[\begin{pmatrix}\phi_{9}\\\phi_{10}\end{pmatrix} \cdot \begin{pmatrix}\phi_{9}\\\phi_{10}\end{pmatrix}\right]
\end{equation}
where we suppressed coefficients.  The bracket structure between the
$\phi_{11}$/$\phi_{12}$  and $\phi_{9}$/$\phi_{10}$ is a consequence of the
non-Abelian $D_4$ symmetry, where these fields transform as a doublet.
The dot `$\,\cdot\,$' indicates the standard scalar product. Note that there are
invariants with more than two $D_4$ charged fields which cannot be written in
terms of a scalar product. Further,
$S$ is the dilaton and the coefficient $a=8\pi^2$ in
$\mathrm{e}^{-a\,S}$ is such that $\mathrm{e}^{-a\,S}$ has positive anomalous
charge with respect to the normalized generator of the `anomalous' \U1. This generator
is chosen such that it is the gauge embedding of the anomalous space group
element\footnote{See \cite{Araki:2008ek} for the discussion in a more general
context. Note that we can always bring the anomalous space group element to the
form $(\theta^{k}\,\omega^\ell,0)$ by redefining the model input appropriately.
This amounts to a redefinition of the `origin' of the orbifold.} (cf.\
\Eqref{eq:anomalous_space_group_element}),
\begin{equation}
 \mathsf{t}_\mathrm{anom}~=~W_3+\E8\times\E8~\text{lattice vectors}\;.
\end{equation}
The discrete Green-Schwarz mechanism is discussed in detail in \cite{LRRRSSV2}.

\paragraph{Solution to the $\boldsymbol{\mu}$ problem.} The $\Z4^R$ anomaly has
important consequences for the MSSM $\mu$ problem. The $\mu$ term is forbidden
perturbatively by $\Z4^R$, however, it appears at the non-perturbative level.
Further, this model shares with the mini-landscape models the property that  any
allowed superpotential term can serve as an effective $\mu$ term (cf.\ the
discussion in \cite{Kappl:2008ie}). This fact can be seen from
higher-dimensional gauge invariance \cite{Brummer:2010fr}. Therefore, the
(non-perturbative) $\mu$ term is of the order of the gravitino mass,
\begin{equation}
 \mu~\sim~\langle\mathscr{W}\rangle~\sim~m_{3/2}
\end{equation}
in Planck units. If some `hidden' sector dynamics induces a
non-trivial $\langle\mathscr{W}\rangle$, the $\mu$ problem is
solved.

In our model, we have only a `toy' hidden sector with an unbroken \SU2 gauge
group and one pair of massless doublets whose mass term is prohibited by
$\Z4^R$. This sector has the structure discussed by Affleck, Dine and Seiberg
(ADS) \cite{Affleck:1983mk}. We find that the ADS superpotential is $\Z4^R$
covariant. However, the hidden gauge group is probably too small for generating
a realistic scale of supersymmetry breakdown. Yet there are alternative ways,
such as the one described in \cite{Kappl:2008ie}, for generating a
hierarchically small $\langle\mathscr{W}\rangle$.

\paragraph{Origin of $\boldsymbol{\Z4^R}$.}
In the orbifold CFT description the $\Z4^R$ originates from the so-called
$H$-momentum selection rules \cite{Dixon:1986qv} (see also
\cite{Kobayashi:2004ya,Buchmuller:2006ik}). These selection rules appear as
discrete $R$ symmetries in the effective field theory description of the model.
We would like to stress that in large parts of the literature the order of these
symmetries was given in an unfortunate way. This criticism also applies to the
papers by some of the authors of this study. For instance, the $\Z2$ orbifold
plane was said to lead to a $\Z2^R$ symmetry, but it turned out that there are
states with half-integer charges. We find it more appropriate to call this
symmetry $\Z4^R$, and to deal with integer charges only. In our model we have
three $\Z4^R$ symmetries at the orbifold point, stemming from the three $\Z2$
orbifold planes.

$H$-momentum corresponds to angular momentum in the compact space; therefore the
discrete $R$ symmetries can be thought of as discrete remnants of the Lorentz
symmetry of internal dimensions. That is to say that the orbifold
compactification breaks the Lorentz group of the tangent space to a discrete
subgroup. In this study we content ourselves with the understanding that these
symmetries appear in the CFT governing the correlators to which we match the
couplings of our effective field theory. The precise geometric interpretation of
this symmetry in field theory will be discussed elsewhere.

The actual $\Z4^R$ charges of $[\SU3\times\SU2\times\SU2]_\mathrm{hid}$
invariant expressions in this model are given by
\begin{equation}\label{eq:Z4Rcharge}
 q_{\Z4^R}~=~q_X+R_2+2n_3\;,
\end{equation}
where $q_X$ is the \U1 charge  generated by
\begin{equation}
\mathsf{t}_X~=~
\left(4, 0, 10, -10, -10, -10, -10, -10\right)\,
\left(-10, 0, 5, 5, -5, 15, -10, 0\right)\;,
\end{equation}
$R_2$ denotes the $R$ charge with respect to the second orbifold plane and $n_3$ is the
localization quantum number in the third torus. The relevant quantum numbers are
given in table~\ref{tab:fullspectrum}. The expression \eqref{eq:Z4Rcharge} for
$q_{\Z4^R}$ is not unique, there are 17 linear combinations of \U1 charges and
discrete quantum numbers which can be used to rewrite the formula without
changing the $\Z4^R$ charges. Also the \U1 factors contained in
$[\SU3\times\SU2\times\SU2]_\mathrm{hid}$ can be used to redefine
$\mathsf{t}_X$. We refrain from spelling this out as we find it more convenient
to work with invariant monomials (cf.\ the discussion in
appendix~\ref{app:SUSYVacuum}). It is straightforward to see that all monomials
we switch on have $R$ charge 0.

\section{Summary}
\label{sec:Summary}

We have re-emphasized the important role of discrete symmetries in string model
building.  As an application, we discussed an explicit string model which
exhibits MSSM vacua with a $\Z4^R$ symmetry, which has recently been shown to be
the unique symmetry for the MSSM that forbids the $\mu$ term at the perturbative
level, allows Yukawa couplings and neutrino masses, and commutes with \SO{10}.
This $\Z4^R$ has a couple of appealing features. First, the
$\mu$ term and dangerous dimension five proton decay operators are forbidden at
the perturbative level and appear only through (highly suppressed)
non-perturbative effects. Second, at the perturbative level, the expectation
value of the superpotential is zero; a non-trivial expectation value is
generated by non-perturbative effects. These two points imply that $\mu$ is of
the order of the gravitino mass $m_{3/2}$, which is set by the expectation value
of the superpotential (in Planck units).

The model is a $\Z2\times\Z2$ orbifold compactification of the $\E8\times\E8$
heterotic string. We discussed how to search for field configurations which
preserve $\Z4^R$ and how to find supersymmetric vacua within such
configurations. The Hilbert basis method allowed us to construct a basis for
all gauge invariant holomorphic monomials, and therefore to survey the
possibilities of satisfying the $D$-term constraints.
As we have seen, in the case of residual $R$ symmetries it may
in principle happen that the $F$-term equations overconstrain the system. We have
explicitly verified that this is not the case in our model, i.e.\ there are
supersymmetric vacua with the exact MSSM spectrum and a residual $\Z4^R$
symmetry. Let us highlight the features of the model:
\begin{itemize}
 \item exact MSSM spectrum, i.e.\ no exotics;
 \item almost all singlet fields/moduli are fixed in a
 supersymmetric way;
 \item non-local GUT breaking, i.e.\ the model is consistent with MSSM precision
 gauge unification;
 \item dimension four proton decay operators are completely absent as $\Z4^R$
 contains the usual matter parity as a subgroup;
 \item dimension five proton decay operators only appear at the non-perturbative
 level and are completely harmless;
 \item the gauge and top-Yukawa couplings coincide at tree level;
 \item see-saw suppressed neutrino masses;
 \item $\mu$ is related to the vacuum expectation value of the superpotential
 and therefore of the order of the gravitino mass;
 \item there is an \SU5 GUT relation between the $\tau$ and bottom masses.
\end{itemize}
There are also two drawbacks: first, there are also \SU5 relations for the light
generations and second the hidden sector gauge group is only \SU2 and therefore
probably too small for explaining an appropriate scale of
dynamical supersymmetry breaking.

Although we have obtained a quite promising string vacuum, the main focus of
this paper was on developing new methods rather than working out the
phenomenology of a model. We have discussed in detail how to determine the
residual discrete symmetries of a given VEV configuration. As a consequence, we
could immediately understand the features of such a configuration. For instance,
in earlier studies
\cite{Buchmuller:2005jr,Buchmuller:2006ik,Lebedev:2006kn,Lebedev:2007hv} we had
to explicitly identify couplings that are consistent with the string selection
rules in order to show that all exotics decouple and the Yukawa couplings have
full rank. This is a very time-consuming task in practice. With the new methods
we could obtain this information by just looking at the remnant symmetries. We
have performed extensive cross-checks in order to show that both methods yield
the same results. We have also shown how to search for vacua with a given
symmetry. Further, we presented the Hibert basis method which allows us to
survey all $D$-flat directions comprised of a selected set of fields in very
short time. It will be interesting to apply, and to extend, our methods to other
examples.

\subsection*{Acknowledgements}

We thank James Gray, Thomas Grimm, Arthur Hebecker, Raymond
Hemmecke, Christoph L\"udeling, Graham Ross and Timo Weigand for
valuable discussions. This research was supported by the DFG cluster
of excellence Origin and Structure of the Universe, the
\mbox{SFB-Transregio} 27 ``Neutrinos and Beyond'', LMUExcellent, the
Graduiertenkolleg ``Particle Physics at the Energy Frontier of New
Phenomena'' by Deutsche Forschungsgemeinschaft (DFG), and by the
National Science Foundation under Grant No.\ PHY05-51164. We would
like to thank the Aspen Center for Physics and the KITP in Santa
Barbara, where some of this work has been carried out, for
hospitality and support.  S.R. acknowledges partial support from DOE
Grant DOE/ER/01545-890.

\appendix
\section{Discrete anomalies}

\subsection{Discrete anomaly calculation}
\label{app:DiscreteAnomalyCalculation}

We calculate the anomaly of the $\Z4^R$ symmetry of the
configuration discussed in section~\ref{sec:the_model}.
The fermions of a superfield with $\Z4^R$ charge $q$ have $\Z4^R$ charge $q-1$ because we work in a convention where the superpotential carries charge two. Only
massless states contribute to the anomaly (cf.\ \cite{Araki:2008ek}).
We can hence limit ourselves to the MSSM field content. Since all matter fields carry $\Z4^R$ charge one, the corresponding fermion is uncharged and does not contribute to the anomaly. Thus, we end up with the following contributions:
\[
\begin{array}{lrc|clr}
  \SU3_C & & & & \SU2_\mathrm{L} & \\
 \hline
  & & & & h_{u,d}  & 2\cdot\frac{1}{2}\cdot 3~=~3 \\
 \text{gauginos} & c_2(\boldsymbol{8})~=~3 & & &
 \text{gauginos}    & c_2(\boldsymbol{3})~=~2\\
    \hline
 & 3 & & & & 5
\end{array}
\]
The factor $1/2$ in the Higgs contributions is the Dynkin index.  The
anomaly condition is that the sum over all charges be equal mod 2. The total contribution in both cases is odd. That is, both symmetries
appear anomalous, and the anomaly is, in particular, universal, as required for
the Green-Schwarz mechanism to work.

\subsection{Anomaly Mixing}

In our model we have two symmetries which appear anomalous. On the one hand, we
have the anomalous $\U1_\text{anom}$. On the other hand, the space group
selection rule $\Z2^{n_3}$ for $n_3$ (cf.\ \cite{Blaszczyk:2009in}),
corresponding to the anomalous space group element in
\Eqref{eq:anomalous_space_group_element}, is anomalous. We wish to answer the
question if it is possible to rotate the $\Z2^{n_3}$ anomaly completely into
$\U1_\text{anom}$.

Consider a setting with $\U1\times\Z{N}$ symmetry which appear anomalous. We
will denote the $\U1$ charge by $Q^{(i)}$ and the \Z{N} charges by $q^{(i)}$,
and we will assume all charges to be integers. Suppose we have a gauge group $G$.
The anomaly coefficients read
\begin{subequations}
\begin{alignat}{2}
 G-G-\U1~:&\quad \sum_f Q^{(f)}\, \ell(\boldsymbol{r}^{(f)})~&=~A \; , \\
 G-G-\Z{N}~:&\quad \sum_f q^{(f)}\, \ell(\boldsymbol{r}^{(f)})~&=~B \; ,
\end{alignat}
\end{subequations}
where the sums run over the irreducible representations of $G$ and
$\ell(\boldsymbol{r})$ is the Dynkin index of the representation
$\boldsymbol{r}$. Specifically, $\ell=1/2$ for the fundamental
representation of \SU{N}.

We can redefine the $\Z{N}$ charges by shifting them by integer multiples of the
\U1 charges. That is, we can define new \Z{N} charges
$q'^{(i)}=q^{(i)}+n\,Q^{(i)}$. Then the new $G-G-\Z{N}$ anomaly coefficient is
given by
\begin{equation}
 \sum_f q'^{(f)} \ell(\boldsymbol{r}^{(f)})
 ~=~
 \sum_f \left(q^{(f)}+n\,Q^{(f)}\right)\, \ell(\boldsymbol{r}^{(f)})
 ~=~B+n\,A
\end{equation}
with $n\in\Z{}$. Anomaly freedom
requires~\cite{Ibanez:1991hv,Araki:2008ek}
\begin{equation}
 \sum_f q'^{(f)} \ell(\boldsymbol{r}^{(f)})
 ~=~ 0 \mod \eta \quad \text{where}\quad
 \eta ~=~ \left\lbrace \begin{array}{ll}N, & N \:\text{odd} \\ \frac{N}{2}, &  N \:\text{even}\end{array} \right. \: .
\end{equation}
Hence, the \Z{N} can be made anomaly-free if there is a solution to
\begin{equation}
 B+n\,A ~=~ 0 \mod \eta
\end{equation}
for $n\in\Z{}$.

The anomaly coefficient for the anomalous $\U1_\text{anom}$ in our model is given by
\begin{equation}
 \SU3_C-\SU3_C-\U1_\mathrm{anom}~:\quad A~=~15
\end{equation}
for the SM gauge group $\SU3_C$. The $\U1_\text{anom}$ charges are normalized to
integers. The anomaly coefficients are the same for $\SU2_\mathrm{L}$ and for
the gauge group factors of the hidden sector because of
the Green-Schwarz mechanism. In addition, the $\Z2^{n_3}$ anomaly coefficient
turns out to be
\begin{equation}
  \SU3_C-\SU3_C-\Z2^{n_3}~:\quad B~=~\frac{1}{2} \:.
\end{equation}
The equation
\begin{equation}
 B+n\,A ~=~ \frac{1}{2}+15\, n ~=~0\mod 1
\end{equation}
has no solution for $n\in\Z{}$ and therefore the anomaly of $\Z2^{n_3}$ cannot
be removed. Altogether we have demonstrated that there are (at least) two
independent anomalies, i.e.\ the imaginary part of the dilaton shifts both under
$\U1_\mathrm{anom}$ and $\Z2^{n_3}$ transformations and there is no way of
rotating the $\Z2^{n_3}$ anomaly into $\U1_\mathrm{anom}$.

\section{Identifying the $\boldsymbol{\Z4^R}$ symmetry}
\label{app:IdentifyingZ4R}

In section~\ref{sec:the_model} we explained our strategy for constructing promising vacuum configurations. A crucial point is the identification of vacua that exhibit a $\Z4^R$ symmetry, with charges given in table~\ref{tab:DiscreteChargesMatterZ2xZ2_1-1}. In this appendix, we will give some details of how this identification is done.

In a particular vacuum, our models exhibit Abelian discrete symmetries which are
calculated by the methods described in \cite{Petersen:2009ip}. A finite Abelian group can always be written as a direct product
$G=H_{p_1}\times\ldots\times H_{p_n}$ where the $p_i$ are pairwise distinct
primes and $H_p=\Z{p^{e_1}}\times\ldots\times\Z{p^{e_m}}$ with $e_1\leq \cdots
\leq e_m$ positive integers. In a first step, we have to make sure that $G$ has a \Z4 subgroup. This can be done by looking at the subgroup $H_2$ of $G$.

In the next step, we want to know whether the \Z4 is of $R$ or non-$R$ type. In order to answer this question, we only have to look at the transformation of the superpotential under the \Z4; thus
one can unambiguously see if the \Z4 is $R$ or non-$R$.

If the considered vacuum exhibits a $\Z4^R$ symmetry, we have to check that the charges of the matter fields match the ones in table~\ref{tab:DiscreteChargesMatterZ2xZ2_1-1}. A technical problem appears if $H_2$ consists of more than one factor, i.e. if there is an additional $\Z{2^n}$ symmetry where $n$ is a positive integer. In this case there are many equivalent charge
assignments some of which will make the $\Z4^R$ obvious while others will
conceal its existence. This freedom corresponds to the automorphisms of $G$
\cite{Hillar:2007} whose number can be large, e.g.\ $\Z{2}\times\Z{4}\times\Z{4}$
has 1536 automorphisms. An important fact is that the automorphism group
factorizes in the way we have written $G$, i.e.\ $\text{Aut}(H_{p_1}\times
H_{p_2})\cong \text{Aut}(H_{p_1})\times \text{Aut}(H_{p_2})$. Thus, we only need
to look at $H_2$.

The automorphisms can be represented by certain matrices acting on charge
vectors \cite{Hillar:2007}. To see whether the $\Z4^R$ is present in a
vacuum, we scan over all possible charge assignments of $H_2$ and look for a
\Z{4} subgroup under which all SM matter fields have charge 1. The other states can be
even or odd, as long as they are vector-like and can be decoupled.

To illustrate this procedure, let us look at a simple example in table
\ref{tab:example_Z4}. The two charge assignments are equivalent. While
in \subref{tab:example_Z4_1} it is not obvious that there is a \Z4 subgroup under which all $\psi$ have charge 1, in the second charge assignment in \subref{tab:example_Z4_2} all fields have charge 1 with respect to the second \Z4 subgroup.

\begin{table}[ht]
\centerline{
\subtable[\label{tab:example_Z4_1}]{
\begin{tabular}{cccc}
\toprule[1.3pt]
& \Z2 &  \Z4 & \Z4\\
$\psi_1$ & 1 & 0 & 3 \\
$\psi_2$ & 1 & 3 & 2\\
$\psi_3$ & 0 & 0 & 3\\
\bottomrule[1.3pt]
\end{tabular}
}\quad
\subtable[\label{tab:example_Z4_2}]{
\begin{tabular}{c ccc}
\toprule[1.3pt]
& \Z2 &  \Z4 & \Z4\\
$\psi_1$ & 0 & 1 & 2\\
$\psi_2$ & 1 & 1 & 3\\
$\psi_3$ & 1 & 1 & 0\\
\bottomrule[1.3pt]
\end{tabular}
}
}
\caption{An example for a hidden \Z4 symmetry under which all fields have charge 1. The two charge assignments are equivalent.}
\label{tab:example_Z4}
\end{table}

\section[Hilbert bases and $D$-flatness]
{Hilbert bases and $\boldsymbol{D}$-flatness}
\label{app:HilbertBasis}

\subsection{General discussion}

In this appendix, a simple method is described that allows us to
analyze $D$-flatness.

It is well known that $D$-flat directions correspond to holomorphic gauge
invariant monomials \cite{Buccella:1982nx,Cleaver:1997jb}. We will present now a
method to compute all holomorphic gauge invariant monomials. Let us look at a
theory with gauge group $\U1^n$. As can be easily seen, a monomial
$\phi_1^{n_1}\phi_2^{n_2}\ldots \phi_k^{n_k}$ is $D$-flat under the
$j^\mathrm{th}$ \U1 factor if (cf.\ e.g.\ \cite[appendix~B,
equation~(B3)]{Lebedev:2007hv})
\begin{equation}
\sum_i q^{(j)}_i n_i~=~0
\end{equation}
where $q^{(j)}_i$ is the charge of the field $\phi_i$ under the
$j^\mathrm{th}$ \U1. The index $i$ runs from 1 to the number of fields $k$
in the monomial and the index $j$ over the number of different \U1
factors. We can rewrite this as a matrix equation,
\begin{equation}
\begin{pmatrix}
 q_1^{(1)}&\ldots&q_k^{(1)}\\
 \vdots&&\vdots\\
q_1^{(n)}&\ldots&q_k^{(n)}
\end{pmatrix}
\cdot
\begin{pmatrix}
n_1\\
\vdots\\
n_k\\
\end{pmatrix}
~=~0\;,
\end{equation}
where $n_i$ counts how often a field occurs in the monomial. The
charges can always be scaled to become integers whereas the field
multiplicity $n_i$ has to be a non-negative integer. Thus, the
condition to have a  $D$-flat monomial is
\begin{equation}
\label{equa:hilbert}
 Q\cdot x~=~0\;,\quad Q\in\mathbbm{Z}_{n\times k}\;,\quad x\in\mathbbm{N}^k\;.
\end{equation}
This is a system of homogeneous linear Diophantine equation over non-negative
integers. Such equations can be solved completely by constructing the
corresponding Hilbert basis (see for example  \cite{720989}). A Hilbert basis
$\mathcal{H}(Q)$ is a complete set of all minimal solutions to equation
\eqref{equa:hilbert}. A solution is called minimal if it is non-trivial and
there exists no smaller solution. Here `smaller' means that, given a solution
$x$, there is no other solution $y\neq x$ with $y_i\leq x_i$ for all $i=1\ldots
k$. Given $\mathcal{H}(Q)$ we can construct all non-negative solutions to
equation \eqref{equa:hilbert} by forming linear combinations of the basis
solutions with non-negative integer coefficients. The Hilbert basis $\mathcal{H}(Q)$ is
therefore a basis for all $D$-flat monomials. In practice the Hilbert basis for
a given matrix $Q$ can be computed with the help of computer algebra packages
like \cite{4ti2}. Such packages can be applied even for rather large matrices.

Let us look at an example from \cite{Luty:1995sd}  with four fields. The charges
of the fields are summarized in the charge matrix
\begin{equation}
 Q~=~\begin{pmatrix}
 2&-2&1&-1\\
 0&1&-1&0
 \end{pmatrix}\;.
\end{equation}
The Hilbert basis $\mathcal{H}(Q)$ is given by the three vectors
\begin{equation}
x~=~\begin{pmatrix}
1\\
0\\
0\\
2
\end{pmatrix}\;,\quad
y~=~\begin{pmatrix}
1\\
2\\
2\\
0
\end{pmatrix}\;,\quad
z~=~\begin{pmatrix}
1\\
1\\
1\\
1
\end{pmatrix}\qquad
 x,y,z~\in~\mathcal{H}(Q)\;.
\end{equation}
All holomorphic gauge invariant monomials
$\phi_1^{n_1}\phi_2^{n_2}\phi_3^{n_3}\phi_4^{n_4}$ can be characterized by
four-vectors $w=(n_1,n_2,n_3,n_4)^T$, which are given by $w=\alpha\, x+\beta\,
y+\gamma\, z$ with $\alpha,\beta,\gamma\in \mathbbm{N}$. We recognize an important
property of the Hilbert basis: while the dimension of $D$-flat directions is 2,
i.e.\ the number of fields minus the number of independent $D$-term constraints,
the length of the Hilbert basis is larger, namely 3. There is one relation
between the Hilbert basis elements,
\begin{equation}\label{eq:HilbertRelation}
 x+y~=~2z\;.
\end{equation}
This is the price one has to pay for being able to express any $D$-flat
direction as an integer linear combination of basis monomials.

\subsection{Hilbert basis for the vacuum discussed in section \ref{sec:the_model}}
\label{app:details_hilbert_basis}

With the Hilbert basis method  we could identify a complete set of $D$-flat
directions composed of $\widetilde{\phi}$ fields in \Eqref{eq:phitildefields}.
As discussed, the key feature of this Hilbert basis is that any gauge invariant
holomorphic monomial can be expressed as product of the basic monomials.  We
obtain 6184 monomials. An example of a monomial, which has negative charge under
the anomalous \U1, is given in \Eqref{eq:hilbert_monomial}.

\section{Details of the supersymmetric vacuum configuration}
\label{app:SUSYVacuum}

\subsection{$\boldsymbol{D}$-flatness}

\paragraph{Hidden SU(2) breaking.} We first look at \SU2. We find that we can switch on $y_{3},\ldots ,y_6$.
This leads to the 6 \SU2 invariant monomials
\begin{equation}
\{\mathscr{M}^{(i)}\}_{\SU2} ~ = ~
\{{y_{3}y_{4}},{y_{3}y_{5}},{y_{3}y_{6}},{y_{4}y_{5}},{y_{4}y_{6}},{y_{5}y_{6}}\}
\;.
\end{equation}
There is one relation between the monomials,
\begin{equation}
(y_{3}y_{4})\, (y_{5}y_{6})\,-(y_{3}y_{5})\, (y_{4}y_{6})\,+(y_{3}y_{6})\, (y_{4}y_{5})
~=~0
 \;,
\end{equation}
such that the number of flat directions is 5, which is consistent with 8
components being switched on whereby 3 directions get eaten by the \SU2 gauge
multiplet.
The monomial $y_1y_2$, which also has $R$ charge 0, vanishes. A possible way
to have the above monomials non-vanishing is to set
\begin{subequations}\label{eq:SU2vacuum}
\begin{equation}
{y_{1}^{[1]}}~=~{y_{1}^{[2]}}~=~{y_{2}^{[1]}}~=~{y_{2}^{[2]}}~=~{y_{3}^{[1]}}~=~{y_{4}^{[2]}}~=~0
\end{equation}
and to have, correspondingly,
\begin{equation}
 y_3^{[2]},~y_4^{[1]},~y_5^{[1,2]},~y_6^{[1,2]}~\ne~0\;.
\end{equation}
\end{subequations}

\paragraph{Hidden SU(3) breaking.}
There are 16 \SU3 invariant composites of hidden \SU3 $\rep{3}$- and
$\crep{3}$-plets with $R$ charge 0, and can, from this perspective, acquire a
VEV, namely
\begin{eqnarray}
 \{\mathscr{M}_0^{(i)}\}_{\SU3}
 & = &
 \{{x_{1}\overline{x}_{1}},{x_{2}\overline{x}_{3}},{x_{2}\overline{x}_{4}},
 {x_{2}\overline{x}_{5}},{\overline{x}_{3}x_{4}},
 {\overline{x}_{3}x_{5}},{x_{4}\overline{x}_{4}},{x_{4}\overline{x}_{5}},
 {x_{5}\overline{x}_{4}},
 {x_{5}\overline{x}_{5}},
 \nonumber\\
 & & {}~~
 {x_{1}x_{2}x_{3}},{x_{1}x_{3}x_{4}},{x_{1}x_{3}x_{5}},
 {\overline{x}_{2}\overline{x}_{3}\overline{x}_{4}},
 {\overline{x}_{2}\overline{x}_{3}\overline{x}_{5}},
 {\overline{x}_{2}\overline{x}_{4}\overline{x}_{5}}\}
\;.
\end{eqnarray}

There are different branches of \SU3 flat directions. In what follows we discuss
one particular of them.

In this branch there are 13 composites with non-zero VEV,
\begin{eqnarray}
\{\mathscr{M}^{(i)}\}_{\SU3} & = &
\{{x_{1}\overline{x}_{1}},{x_{2}\overline{x}_{3}},{x_{2}\overline{x}_{4}},{x_{2}\overline{x}_{5}},{\overline{x}_{3}x_{4}},{\overline{x}_{3}x_{5}},{x_{4}\overline{x}_{4}},{x_{4}\overline{x}_{5}},{x_{5}\overline{x}_{4}},{x_{5}\overline{x}_{5}},
\nonumber\\
& & ~{x_{1}x_{2}x_{3}},{x_{1}x_{3}x_{4}},{x_{1}x_{3}x_{5}}\}
\;,
\end{eqnarray}
while the other 3 monomials vanish,
\begin{subequations}\label{eq:SU3monomialRelations2}
\begin{equation}
(\overline{x}_{2}\overline{x}_{4}\overline{x}_{5})
~=~
(\overline{x}_{2}\overline{x}_{3}\overline{x}_{5})
~=~
(\overline{x}_{2}\overline{x}_{3}\overline{x}_{4}) ~ = ~ 0 \;.
\end{equation}
Assuming that the above $\{\mathscr{M}^{(i)}\}_{\SU3}$ VEVs do not vanish,
we arrive at the relations
\begin{align}
(\overline{x}_{3}x_{4}) & = ~ \frac{(\overline{x}_{3}x_{5})\,(x_{1}x_{3}x_{4})}{(x_{1}x_{3}x_{5})}
\;, &
(x_{2}\overline{x}_{3}) & = ~ -\frac{(\overline{x}_{3}x_{5})\,(x_{1}x_{2}x_{3})}{(x_{1}x_{3}x_{5})}
\;,\\
(x_{2}\overline{x}_{5}) & = ~ -\frac{(x_{1}x_{2}x_{3})\,(x_{4}\overline{x}_{5})}{(x_{1}x_{3}x_{4})}
\;, &
(x_{5}\overline{x}_{5}) & = ~ \frac{(x_{1}x_{3}x_{5})\,(x_{4}\overline{x}_{5})}{(x_{1}x_{3}x_{4})}
\;,\\
(x_{2}\overline{x}_{4}) & = ~ -\frac{(x_{1}x_{2}x_{3})\,(x_{4}\overline{x}_{4})}{(x_{1}x_{3}x_{4})}
\;, &
(x_{5}\overline{x}_{4}) & = ~ \frac{(x_{1}x_{3}x_{5})\,(x_{4}\overline{x}_{4})}{(x_{1}x_{3}x_{4})}
\;.
\end{align}
\end{subequations}
This leaves us with 7 independent \SU3 monomials, a possible choice is given by
\begin{equation}
\{{\overline{x}_{3}x_{5}},{x_{1}\overline{x}_{1}},{x_{4}\overline{x}_{4}},{x_{4}\overline{x}_{5}},{x_{1}x_{2}x_{3}},{x_{1}x_{3}x_{4}},{x_{1}x_{3}x_{5}}\}
\;.
\end{equation}
So we see explicitly that this branch of $D$-flat directions has dimension 7.
This is in agreement with the result obtained with the STRINGVACUA package
\cite{Gray:2008zs}.

We can satisfy the constraints by setting (in an appropriate gauge) various
components to zero. The only non-vanishing components are
\begin{equation}\label{eq:SU3vacuum}
 x_1^{[3]},~x_2^{[2]},~x_3^{[1]},~x_4^{[2]},~x_5^{[2]},
~\overline{x}_1^{[3]},~\overline{x}_3^{[2]}
,~\overline{x}_4^{[2]},~\overline{x}_5^{[2]}
~\ne~0\;.
\end{equation}

\paragraph{Abelian singlets.}
We can switch on the Abelian singlets
\[
 \{\mathscr{M}^{(i)}_0\}_\mathrm{singlet}
 ~=~
 \{{\phi_{1}},{\phi_{2}},{\phi_{3}},{\phi_{4}},{\phi_{5}},{\phi_{6}},{\phi_{7}},{\phi_{8}},
 {\phi_{9}},{\phi_{10}},{\phi_{11}},{\phi_{12}},{\phi_{13}},{\phi_{14}}\}
\;.\]
Since there are 8 \U1 factors that get broken, there are 8 additional $D$-term
constraints.

\paragraph{Cancellation of the FI term.} With the Hilbert basis method we were
able to compute all gauge invariant monomials carrying negative charge with repect to
$\U1_\mathrm{anom}$. An example is
\begin{equation}
 \mathscr{M}_\mathrm{FI}
 ~=~\phi_{11}^4\,\phi_{4}\,\phi_{7}^2\,\phi_{8}\,\phi_{9}^2\;. \label{eq:hilbert_monomial}
\end{equation}

\subsection{$\boldsymbol{F}$-flatness}

\paragraph{Remnant $\boldsymbol{\Z4^R}$ symmetry.}
Switching on the above fields breaks the gauge, $R$ and other discrete
symmetries down to $G_\mathrm{SM}\times\Z4^R\times[\SU2]$.
We decompose the moduli space in \SU3 and \SU2 composites and basic fields
$\mathscr{M}_r^{(m)}$ where $r$ denotes the $\Z4^R$ charge of the corresponding
objects.
A prominent role will be played by the singlet fields with $R$ charge 2, which
are given by
\begin{eqnarray}
\{\mathscr{M}_2^{(i)}\}_{\SU3}
& = &
\{{x_{1}\overline{x}_{3}},{x_{1}\overline{x}_{4}},{x_{1}\overline{x}_{5}},
{\overline{x}_{1}x_{2}},{\overline{x}_{1}x_{4}},{\overline{x}_{1}x_{5}},
{\overline{x}_{2}x_{3}},
\nonumber\\
& &{} ~~
{x_{2}x_{3}x_{4}},{x_{2}x_{3}x_{5}},{x_{3}x_{4}x_{5}},
{\overline{x}_{1}\overline{x}_{2}\overline{x}_{3}},{\overline{x}_{1}\overline{x}_{2}\overline{x}_{4}},{\overline{x}_{1}\overline{x}_{2}\overline{x}_{5}}
\}
\;,\nonumber\\
\{\mathscr{M}_2^{(i)}\}_{\SU2}
& = &
\{{y_{1}y_{3}},{y_{1}y_{4}},{y_{1}y_{5}},{y_{1}y_{6}},{y_{2}y_{3}},{y_{2}y_{4}},{y_{2}y_{5}},{y_{2}y_{6}}\}
\;,\nonumber\\
\{\mathscr{M}_2^{(i)}\}_{\mathrm{singlet}}
& = &
\{{{{\overline{\phi}_{1}},{\overline{\phi}_{2}},{\overline{\phi}_{3}},\overline{\phi}_{4}},{\overline{\phi}_{5}},{\overline{\phi}_{6}},\overline{\phi}_{7}},{\overline{\phi}_{8}},{\overline{\phi}_{9}},{\overline{\phi}_{10}},{\overline{\phi}_{11}},{\overline{\phi}_{12}}\}
\;.
\end{eqnarray}

\paragraph{$\boldsymbol{F}$-term constraints.} One can use the above monomials
for counting the independent $F$-term constraints. As discussed in
section~\ref{sec:GeneralPicture}, the superpotential will be of
the form
\begin{equation}
 \mathscr{W}~=~\sum_m \mathscr{M}_2^{(m)}\cdot
    f_2^{(m)}\left(\mathscr{M}_0^{(1)},\dots\right) + \dots\;,
\end{equation}
where the omission contains only terms which are at least quadratic in
$\mathscr{M}_{\ge1}^{(m)}$, and the $f_2^{(m)}$ are some functions of the
monomials with $R$ charge 0. The potentially non-trivial $F$-terms are then
\begin{equation}
 \left.\frac{\partial\mathscr{W}}{\partial\phi_i}
 \right|_{\phi_i~=~\langle\phi_i\rangle}
 ~=~
 \sum_m\left.\frac{\partial\mathscr{M}_2^{(m)}}{\partial\phi_i}
 \cdot f_2^{(m)}\left(\mathscr{M}_0^{(1)},\dots\right)\right|_{\phi_i~=~\langle\phi_i\rangle}
\end{equation}
as we look at vacua with unbroken $\Z4^R$, i.e.\
$\mathscr{M}_{\ge1}^{(m)}=0$.
For $\mathscr{M}_2^{(m)}\in\{\mathscr{M}_2^{(i)}\}_{\mathrm{singlet}}$ each
equation gives a non-trivial constraint on the $\mathscr{M}_0^{(j)}$.
The number of independent constraints
is given by the rank of the matrix
\begin{equation}\label{eq:N-Matrix}
 \mathscr{N}~=~(\mathscr{N}_{ij})~=~\frac{\partial\mathscr{M}_2^{(i)}}{\partial\phi^{(j)}}\;,
\end{equation}
where the $\phi^{(j)}$ comprise all component fields appearing in monomials,
evaluated at the vacuum.

We evaluated the rank of the $\mathscr{N}$ matrix for the \SU3 and \SU2
monomials in the vacuum defined by \eqref{eq:SU2vacuum} and \eqref{eq:SU3vacuum}.
The result is that there are 7 independent $F$-term constraints in the \SU3 case
and 4 in the \SU2 case.
Adding the constraints from the non-Abelian singlets
with $R$ charge 2, we therefore obtain $7+4+12=23$ $F$-term conditions.
At this point, the supersymmetry conditions
seem to over-constrain the system, as the number of $D$-flat directions is
$7+5+14-8=18$. Note, however, that there are 6 additional
degrees of freedom which we have not discussed yet: the $T_i$- and $U_i$-moduli
of our $\Z2\times\Z2$ orbifold. The functions $f_2^{(m)}$ will also depend on
these fields, which obviously have $R$ charge 0 (cf.\ the discussion in
\cite{Brummer:2010fr}). Using these additional degrees of freedom we will
\emph{generically} be able to satisfy the constraints, and \emph{generically}
there will be only one flat direction in the moduli space formed out of the
standard model singlet degrees of freedom!\footnote{Of course,
there is the hidden \SU2 sector which contains further massless degrees of
freedom. We kept this \SU2 unbroken on purpose as it may serve as a toy hidden
sector for dynamical supersymmetry breakdown.}

There are also \SU3 invariant composites with odd $\Z4^R$ charge. The
superpotential will contain terms of the form
\begin{equation}
 \mathscr{W}~\supset~\sum_{m,n} \mathscr{M}_1^{(m)}\cdot\mathscr{M}_1^{(n)}\cdot
    f_1^{(m,n)}\left(\mathscr{M}_0^{(1)},\dots\right)
\end{equation}
and analogous terms for the $\mathscr{M}_3^{(m)}$. This will then lead to
non-trivial mass terms for the vanishing \SU3 triplets and anti-triplets.
Analogous statements hold for the other fields with odd $R$ charges.

In summary, we expect that the vacuum discussed here is such that supersymmetry
conditions can be satisfied. Moreover, we find that all but one of the fields are fixed
by the $D$- and $F$-term constraints.  Unlike in the case without a residual
$\Z4^R$ symmetry, due to the $\Z4^R$ the superpotential expectation value is
guaranteed to vanish at the perturbative level.

\section{Details of the model}
\label{app:details_model}

The orbifold model is defined by a torus lattice that is spanned by six orthogonal vectors $e_\alpha$, $\alpha=1,\ldots, 6$, the $\Z2\times\Z2$ twist vectors $v_1=(0,1/2,-1/2)$ and $v_2=(-1/2,0,1/2)$, and the associated shifts
\begin{subequations}
\begin{eqnarray}
V_1 & = & \left(-\frac{1}{2},-\frac{1}{2},           0, 0, 0, 0, 0, 0\right) \left( 0, 0, 0, 0, 0, 0, 0, 0\right) \: , \\
V_2 & = & \left(           0, \frac{1}{2},-\frac{1}{2}, 0, 0, 0, 0, 0\right) \left( 0, 0, 0, 0, 0, 0, 0, 0\right) \: ,
\end{eqnarray}
\end{subequations}
and the six discrete Wilson lines
\begin{subequations}
\begin{eqnarray}
W_1 & = & \left( 0^8 \right) \left( 0^8 \right)\: , \\
W_3 & = & \left( \frac{3}{2}, \frac{1}{2}, \frac{1}{2}, \frac{1}{2}, \frac{1}{2}, \frac{1}{2}, \frac{1}{2},-\frac{1}{2} \right)
          \left( 0, 0, \frac{1}{2}, \frac{1}{2}, \frac{1}{2}, \frac{1}{2}, 1, 1\right)\: , \\
W_5 & = & \left(-\frac{7}{4}, \frac{7}{4},-\frac{1}{4},-\frac{3}{4}, \frac{1}{4}, \frac{1}{4}, \frac{1}{4},-\frac{3}{4} \right)
          \left(-\frac{3}{4}, \frac{5}{4},-\frac{5}{4},-\frac{5}{4}, \frac{1}{4}, \frac{1}{4},-\frac{3}{4}, \frac{5}{4} \right)\: , \\
W_6 & = & \left( \frac{3}{2}, \frac{1}{2},-\frac{3}{2},-\frac{1}{2},-\frac{1}{2},-\frac{3}{2}, \frac{1}{2}, \frac{1}{2} \right)
          \left(-\frac{3}{2},-\frac{1}{2},-\frac{1}{2}, \frac{3}{2},-\frac{3}{2},-\frac{1}{2},-\frac{3}{2}, \frac{3}{2} \right) \: , \\
W_2 &=& W_4 ~ = ~ W_6\; ,
\end{eqnarray}
\end{subequations}
corresponding to the six torus directions $e_\alpha$. Additionally, we divide out the \Z2 symmetry corresponding to
\begin{equation}
 \tau ~=~ \frac{1}{2} (e_2 + e_4 + e_6)
\end{equation}
with a gauge embedding denoted by $W$ (the freely acting Wilson line) where
\begin{equation}
 W~=~ \frac{1}{2} (W_2 + W_4 + W_6) ~=~ \frac{3}{2} W_2 \;.
\end{equation}
The anomalous space group element reads
\begin{equation}
g_\text{anom} ~=~ (k,\ell;n_1,n_2,n_3,n_4,n_5,n_6)
 ~=~ (0, 0;0, 0, 1, 0, 0, 0)\:, \label{eq:anomalous_space_group_element}
\end{equation}
where the boundary conditions of twisted string are
\begin{equation}
 X(\tau,\sigma+2\pi)
 ~=~
 \vartheta^k\,\omega^\ell\,X(\tau,\sigma)+ n_\alpha e_\alpha
\end{equation}
with $\vartheta$ and $\omega$ denoting the rotations corresponding to $v_1$ and
$v_2$. The spectrum is given in table~\ref{tab:states_labels}. In addtion there
are 37 $G_\mathrm{SM}\times[\SU3\times\SU2\times\SU2]_\mathrm{hid}$ singlets.
\begin{table}[htb]
\centering
\begin{tabular}{cccccccccccc}
\toprule[1.3pt]
Label & $q_i$ & $\bar u_i$ & $\bar D_i$ & $D_i$ & $L_i$ & $\bar L_i$ & $\bar e_i$ & $x_i$ & $\bar x_i$ & $y_i$ & $z_i$ \\ \midrule
\# & 3 & 3 & 6 & 3 & 9 & 6 & 3 & 5 & 5 & 6 & 6\\ \midrule
$\SU3_C$ & $\rep{3}$ & $\crep{3}$ & $\crep{3}$ & $\rep{3}$ & $\rep{1}$ & $\rep{1}$ & $\rep{1}$ & $\rep{1}$ & $\rep{1}$ & $\rep{1}$ & $\rep{1}$\\
 $\SU2_\mathrm{L}$ & $\rep{2}$ & $\rep{1}$ & $\rep{1}$ & $\rep{1}$ & $\rep{2}$ & $\rep{2}$ & $\rep{1}$ & $\rep{1}$ & $\rep{1}$ & $\rep{1}$ & $\rep{1}$\\
 $\U1_Y$ & $\tfrac{1}{6}$ & -$\tfrac{2}{3}$ & $\tfrac{1}{3}$ & -$\tfrac{1}{3}$ & -$\tfrac{1}{2}$ & $\tfrac{1}{2}$ & 1 & 0 & 0 & 0 & 0\\ \midrule
 $\SU3$ & $\rep{1}$ & $\rep{1}$ & $\rep{1}$ & $\rep{1}$ & $\rep{1}$ & $\rep{1}$ & $\rep{1}$ & $\rep{3}$ & $\crep{3}$ & $\rep{1}$ & $\rep{1}$\\
 $\SU2$ & $\rep{1}$ & $\rep{1}$ & $\rep{1}$ & $\rep{1}$ & $\rep{1}$ & $\rep{1}$ & $\rep{1}$ & $\rep{1}$ & $\rep{1}$ & $\rep{2}$ & $\rep{1}$\\
 $\SU2$ & $\rep{1}$ & $\rep{1}$ & $\rep{1}$ & $\rep{1}$ & $\rep{1}$ & $\rep{1}$ & $\rep{1}$ & $\rep{1}$ & $\rep{1}$ & $\rep{1}$ & $\rep{2}$\\
\bottomrule[1.3pt]
\end{tabular}
\caption{The states with their quantum number w.r.t.\ the SM and the hidden sector.}
\label{tab:states_labels}
\end{table}
In table~\ref{tab:fullspectrum} we list the full spectrum. In addition to the
states shown there, the spectrum contains the following (untwisted) moduli:  the
dilaton $S$, three K\"ahler moduli $T_i$ and three complex structure moduli
$U_i$.

{\scriptsize
\begin{longtable}{|@{\hspace{0.1cm}}l|@{\hspace{0.1cm}}l|@{\hspace{0.1cm}}c@{\hspace{0.1cm}}c@{\hspace{0.15cm}}c@{\hspace{0.25cm}}c@{\hspace{0.25cm}}c@{\hspace{0.25cm}}c@{\hspace{0.25cm}}c@{\hspace{0.25cm}}c@{\hspace{0.25cm}}c|@{\hspace{0.1cm}}c|@{\hspace{0.1cm}}c@{\hspace{0.2cm}}c|}
 \caption{Spectrum of the model at the orbifold point. The last two columns list
 the (g)eneral and the (c)onfiguration labels. If there are two labels in one
 line, this corresponds to the twist parameter $n_1=0$ for the first label and
 $n_1=1$ for the second.  The two states form a doublet under a $D_4$ symmetry.
 In this model the three $\Z4^R$ charges (corresponding to the three $\Z2$ orbifold planes) of the respective sectors read:
 $R(U_1)=(2,0,0)$,  $R(U_2)=(0,2,0)$, $R(U_3)=(0,0,2)$,
 $R(T_{(1,0)})=(0,1,1)$,  $R(T_{(0,1)})=(1,0,1)$ and
 $R(T_{(1,1)})=(1,1,0)$.}
 \label{tab:fullspectrum}
\\
\hline
sector & irrep & $q_\text{anom}$ & $q_{Y}$ & $q_{X}$ & $q_{1}$ & $q_{2}$ & $q_{3}$ & $q_{4}$ & $q_{5}$ & $q_{6}$ & $q_{\Z4^R}$ & (g) & (c) \\
\hline\hline
\endfirsthead
\hline
sector & irrep & $q_\text{anom}$ & $q_{Y}$ & $q_{X}$ & $q_{1}$ & $q_{2}$ & $q_{3}$ & $q_{4}$ & $q_{5}$ & $q_{6}$ & $q_{\Z4^R}$ & (g) & (c) \\
\hline\hline
\endhead
\hline
\endfoot
$U_1$  & $\left(\rep{1},\rep{1},\rep{1},\rep{1},\rep{1}\right)$ & $4$ & $0$ & $-4$ & $72$ & $-88$ & $356$ & $188$ & $-444$ & $60$ & $-4$ & $N_{1}$ & $\phi_{1}$ \\
 & $\left(\rep{1},\rep{2},\rep{1},\rep{1},\rep{1}\right)$ & $-4$ & $-\tfrac{1}{2}$ & $0$ & $-4$ & $0$ & $8$ & $0$ & $0$ & $0$ & $0$ & $\bar{L}_{1}$ & $\bar{h}_{1}$ \\
 & $\left(\rep{1},\rep{2},\rep{1},\rep{1},\rep{1}\right)$ & $4$ & $\tfrac{1}{2}$ & $0$ & $4$ & $0$ & $-8$ & $0$ & $0$ & $0$ & $0$ & $L_{1}$ & $h_{1}$ \\
 & $\left(\rep{1},\rep{1},\rep{1},\rep{1},\rep{1}\right)$ & $-4$ & $0$ & $4$ & $-72$ & $88$ & $-356$ & $-188$ & $444$ & $-60$ & $4$ & $N_{2}$ & $\phi_{2}$ \\
$U_2$  & $\left(\rep{1},\rep{2},\rep{1},\rep{1},\rep{1}\right)$ & $6$ & $\tfrac{1}{2}$ & $-14$ & $10$ & $-14$ & $62$ & $30$ & $-70$ & $10$ & $-12$ & $L_{3}$ & $h_{3}$ \\
 & $\left(\rep{1},\rep{1},\rep{1},\rep{1},\rep{1}\right)$ & $2$ & $0$ & $10$ & $66$ & $-74$ & $286$ & $158$ & $-374$ & $50$ & $12$ & $N_{5}$ & $\phi_{6}$ \\
 & $\left(\rep{1},\rep{1},\rep{1},\rep{1},\rep{1}\right)$ & $-2$ & $0$ & $-10$ & $-66$ & $74$ & $-286$ & $-158$ & $374$ & $-50$ & $-8$ & $N_{6}$ & $\phi_{7}$ \\
 & $\left(\rep{1},\rep{2},\rep{1},\rep{1},\rep{1}\right)$ & $-6$ & $-\tfrac{1}{2}$ & $14$ & $-10$ & $14$ & $-62$ & $-30$ & $70$ & $-10$ & $16$ & $\bar{L}_{3}$ & $\bar{h}_{3}$ \\
$U_3$  & $\left(\rep{1},\rep{1},\rep{1},\rep{1},\rep{1}\right)$ & $2$ & $0$ & $-14$ & $6$ & $-14$ & $70$ & $30$ & $-70$ & $10$ & $-14$ & $N_{3}$ & $\bar\phi_{4}$ \\
 & $\left(\rep{1},\rep{2},\rep{1},\rep{1},\rep{1}\right)$ & $-2$ & $-\tfrac{1}{2}$ & $10$ & $62$ & $-74$ & $294$ & $158$ & $-374$ & $50$ & $10$ & $\bar{L}_{2}$ & $\bar{h}_{2}$ \\
 & $\left(\rep{1},\rep{2},\rep{1},\rep{1},\rep{1}\right)$ & $2$ & $\tfrac{1}{2}$ & $-10$ & $-62$ & $74$ & $-294$ & $-158$ & $374$ & $-50$ & $-10$ & $L_{2}$ & $h_{2}$ \\
 & $\left(\rep{1},\rep{1},\rep{1},\rep{1},\rep{1}\right)$ & $-2$ & $0$ & $14$ & $-6$ & $14$ & $-70$ & $-30$ & $70$ & $-10$ & $14$ & $N_{4}$ & $\bar\phi_{7}$ \\
\hline
\hline
$T_{(1, 0)}^{(*, *, 0, 0, 0, 0)}$  & $\left(\crep{3},\rep{1},\rep{1},\rep{1},\rep{1}\right)$ & $0$ & $-\tfrac{1}{3}$ & $20$ & $2$ & $-2$ & $12$ & $0$ & $0$ & $0$ & $21$ & $\bar{D}_{1}$ & $\bar{d}_{3}$ \\
 & $\left(\rep{1},\rep{2},\rep{1},\rep{1},\rep{1}\right)$ & $0$ & $\tfrac{1}{2}$ & $20$ & $2$ & $-2$ & $12$ & $0$ & $0$ & $0$ & $21$ & $L_{4}$ & $\ell_{3}$ \\
 & $\left(\rep{1},\rep{1},\rep{1},\rep{1},\rep{1}\right)$ & $4$ & $0$ & $-20$ & $2$ & $2$ & $-20$ & $0$ & $0$ & $0$ & $-19$ & $N_{7}$ & $n_{9}$ \\
 & $\left(\rep{1},\rep{1},\rep{1},\rep{1},\rep{1}\right)$ & $2$ & $-1$ & $0$ & $2$ & $0$ & $-4$ & $0$ & $0$ & $0$ & $1$ & $\bar{E}_{1}$ & $\bar{e}_{3}$ \\
 & $\left(\rep{3},\rep{2},\rep{1},\rep{1},\rep{1}\right)$ & $2$ & $-\tfrac{1}{6}$ & $0$ & $2$ & $0$ & $-4$ & $0$ & $0$ & $0$ & $1$ & $Q_{1}$ & $q_{3}$ \\
 & $\left(\crep{3},\rep{1},\rep{1},\rep{1},\rep{1}\right)$ & $2$ & $\tfrac{2}{3}$ & $0$ & $2$ & $0$ & $-4$ & $0$ & $0$ & $0$ & $1$ & $\bar{U}_{1}$ & $\bar{u}_{3}$ \\
$T_{(1, 0)}^{(*, *, 0, 0, 1, 0)}$  & $\left(\rep{1},\rep{1},\rep{1},\rep{1},\rep{1}\right)$ & $0$ & $0$ & $7$ & $24$ & $-31$ & $140$ & $60$ & $-144$ & $20$ & $8$ & $N_{8}$ & $\phi_{8}$ \\
 & $\left(\rep{1},\rep{1},\rep{1},\rep{1},\rep{1}\right)$ & $4$ & $0$ & $-11$ & $48$ & $-57$ & $216$ & $128$ & $-300$ & $40$ & $-10$ & $N_{9}$ & $\bar\phi_{12}$ \\
 & $\left(\rep{1},\rep{2},\rep{1},\rep{1},\rep{1}\right)$ & $4$ & $-\tfrac{1}{2}$ & $-1$ & $-14$ & $15$ & $-62$ & $-30$ & $74$ & $-10$ & $0$ & $\bar{L}_{4}$ & $\bar{h}_{4}$ \\
 & $\left(\rep{3},\rep{1},\rep{1},\rep{1},\rep{1}\right)$ & $4$ & $\tfrac{1}{3}$ & $-1$ & $-14$ & $15$ & $-62$ & $-30$ & $74$ & $-10$ & $0$ & $D_{1}$ & $\delta_{1}$ \\
 & $\left(\rep{1},\rep{1},\rep{1},\rep{1},\rep{1}\right)$ & $-8$ & $0$ & $1$ & $10$ & $-15$ & $70$ & $30$ & $-74$ & $10$ & $2$ & $N_{10}$ & $\bar\phi_{1}$ \\
$T_{(1, 0)}^{(*, *, 0, 0, 1, 1)}$  & $\left(\rep{1},\rep{1},\rep{3},\rep{1},\rep{1}\right)$ & $4$ & $0$ & $-16$ & $-14$ & $16$ & $-62$ & $-34$ & $74$ & $-10$ & $-15$ & $N_{11}$ & $x_{1}$ \\
 & $\left(\rep{1},\rep{1},\rep{1},\rep{1},\rep{1}\right)$ & $4$ & $0$ & $4$ & $-14$ & $16$ & $-70$ & $-26$ & $66$ & $-10$ & $5$ & $N_{12}$ & $n_{1}$ \\
 & $\left(\rep{1},\rep{1},\rep{1},\rep{2},\rep{1}\right)$ & $3$ & $0$ & $-11$ & $-14$ & $16$ & $-70$ & $-30$ & $74$ & $-8$ & $-10$ & $N_{13}$ & $z_{1}$ \\
 & $\left(\rep{1},\rep{1},\rep{1},\rep{1},\rep{2}\right)$ & $3$ & $0$ & $-21$ & $-14$ & $16$ & $-70$ & $-30$ & $82$ & $-12$ & $-20$ & $N_{14}$ & $y_{1}$ \\
$T_{(1, 0)}^{(*, *, 1, 0, 0, 0)}$  & $\left(\rep{1},\rep{1},\crep{3},\rep{1},\rep{1}\right)$ & $2$ & $0$ & $-12$ & $36$ & $-44$ & $182$ & $90$ & $-218$ & $30$ & $-9$ & $N_{15}$ & $\bar{x}_{1}$ \\
 & $\left(\rep{1},\rep{1},\rep{3},\rep{1},\rep{1}\right)$ & $2$ & $0$ & $8$ & $36$ & $-44$ & $174$ & $98$ & $-226$ & $30$ & $11$ & $N_{16}$ & $x_{2}$ \\
 & $\left(\rep{1},\rep{1},\rep{1},\rep{1},\rep{1}\right)$ & $0$ & $0$ & $-2$ & $36$ & $-44$ & $190$ & $90$ & $-226$ & $30$ & $1$ & $N_{17}$ & $n_{2}$ \\
 & $\left(\rep{1},\rep{1},\rep{1},\rep{1},\rep{1}\right)$ & $4$ & $0$ & $-2$ & $36$ & $-44$ & $166$ & $98$ & $-218$ & $30$ & $1$ & $N_{18}$ & $n_{3}$ \\
$T_{(1, 0)}^{(*, *, 1, 0, 0, 1)}$  & $\left(\rep{1},\rep{1},\crep{3},\rep{1},\rep{1}\right)$ & $2$ & $0$ & $3$ & $-26$ & $29$ & $-104$ & $-64$ & $148$ & $-20$ & $6$ & $N_{19}$ & $\bar{x}_{2}$ \\
 & $\left(\rep{1},\rep{1},\rep{1},\rep{1},\rep{1}\right)$ & $2$ & $0$ & $-17$ & $-26$ & $29$ & $-96$ & $-72$ & $156$ & $-20$ & $-14$ & $N_{20}$ & $\bar\phi_{2}$ \\
 & $\left(\rep{1},\rep{1},\rep{3},\rep{1},\rep{1}\right)$ & $4$ & $0$ & $-7$ & $-26$ & $29$ & $-112$ & $-64$ & $156$ & $-20$ & $-4$ & $N_{21}$ & $x_{3}$ \\
 & $\left(\rep{1},\rep{1},\rep{1},\rep{1},\rep{1}\right)$ & $4$ & $0$ & $13$ & $-26$ & $29$ & $-120$ & $-56$ & $148$ & $-20$ & $16$ & $N_{22}$ & $\phi_{3}$ \\
$T_{(1, 0)}^{(*, *, 1, 0, 1, 0)}$  & $\left(\rep{1},\rep{1},\rep{1},\rep{1},\rep{1}\right)$ & $4$ & $0$ & $-11$ & $-10$ & $15$ & $-70$ & $-30$ & $82$ & $-10$ & $-8$ & $N_{23}$ & $\phi_{4}$ \\
 & $\left(\crep{3},\rep{1},\rep{1},\rep{1},\rep{1}\right)$ & $0$ & $-\tfrac{1}{3}$ & $11$ & $14$ & $-15$ & $62$ & $30$ & $-82$ & $10$ & $14$ & $\bar{D}_{2}$ & $\bar\delta_{1}$ \\
 & $\left(\rep{1},\rep{2},\rep{1},\rep{1},\rep{1}\right)$ & $0$ & $\tfrac{1}{2}$ & $11$ & $14$ & $-15$ & $62$ & $30$ & $-82$ & $10$ & $14$ & $L_{5}$ & $h_{4}$ \\
 & $\left(\rep{1},\rep{1},\rep{1},\rep{1},\rep{1}\right)$ & $0$ & $0$ & $21$ & $-48$ & $57$ & $-216$ & $-128$ & $292$ & $-40$ & $24$ & $N_{24}$ & $\phi_{5}$ \\
 & $\left(\rep{1},\rep{1},\rep{1},\rep{1},\rep{1}\right)$ & $-4$ & $0$ & $-17$ & $-24$ & $31$ & $-140$ & $-60$ & $152$ & $-20$ & $-14$ & $N_{25}$ & $\bar\phi_{3}$ \\
$T_{(1, 0)}^{(*, *, 1, 0, 1, 1)}$  & $\left(\rep{1},\rep{1},\rep{1},\rep{1},\rep{2}\right)$ & $3$ & $0$ & $-11$ & $-14$ & $16$ & $-70$ & $-30$ & $74$ & $-8$ & $-8$ & $N_{26}$ & $y_{2}$ \\
 & $\left(\rep{1},\rep{1},\rep{1},\rep{2},\rep{1}\right)$ & $3$ & $0$ & $-21$ & $-14$ & $16$ & $-70$ & $-30$ & $82$ & $-12$ & $-18$ & $N_{27}$ & $z_{2}$ \\
 & $\left(\rep{1},\rep{1},\crep{3},\rep{1},\rep{1}\right)$ & $0$ & $0$ & $26$ & $14$ & $-16$ & $62$ & $34$ & $-82$ & $10$ & $29$ & $N_{28}$ & $\bar{x}_{3}$ \\
 & $\left(\rep{1},\rep{1},\rep{1},\rep{1},\rep{1}\right)$ & $0$ & $0$ & $6$ & $14$ & $-16$ & $70$ & $26$ & $-74$ & $10$ & $9$ & $N_{29}$ & $n_{4}$ \\
\hline
$T_{(0, 1)}^{(n_1, 0, *, *, 0, 0)}$  & $\left(\crep{3},\rep{1},\rep{1},\rep{1},\rep{1}\right)$ & $1$ & $-\tfrac{1}{3}$ & $13$ & $5$ & $-9$ & $47$ & $15$ & $-35$ & $5$ & $13$ & $\bar{D}_{3}$, $\bar{D}_{4}$ & $\bar{d}_{2}$, $\bar{d}_{1}$ \\
 & $\left(\rep{1},\rep{2},\rep{1},\rep{1},\rep{1}\right)$ & $1$ & $\tfrac{1}{2}$ & $13$ & $5$ & $-9$ & $47$ & $15$ & $-35$ & $5$ & $13$ & $L_{6}$, $L_{7}$ & $\ell_{2}$, $\ell_{1}$ \\
 & $\left(\rep{1},\rep{1},\rep{1},\rep{1},\rep{1}\right)$ & $5$ & $0$ & $-27$ & $5$ & $-5$ & $15$ & $15$ & $-35$ & $5$ & $-27$ & $N_{30}$, $N_{36}$ & $n_{5}$, $n_{6}$ \\
 & $\left(\rep{1},\rep{1},\rep{1},\rep{1},\rep{1}\right)$ & $3$ & $-1$ & $-7$ & $5$ & $-7$ & $31$ & $15$ & $-35$ & $5$ & $-7$ & $\bar{E}_{2}$, $\bar{E}_{3}$ & $\bar{e}_{2}$, $\bar{e}_{1}$ \\
 & $\left(\rep{3},\rep{2},\rep{1},\rep{1},\rep{1}\right)$ & $3$ & $-\tfrac{1}{6}$ & $-7$ & $5$ & $-7$ & $31$ & $15$ & $-35$ & $5$ & $-7$ & $Q_{2}$, $Q_{3}$ & $q_{2}$, $q_{1}$ \\
 & $\left(\crep{3},\rep{1},\rep{1},\rep{1},\rep{1}\right)$ & $3$ & $\tfrac{2}{3}$ & $-7$ & $5$ & $-7$ & $31$ & $15$ & $-35$ & $5$ & $-7$ & $\bar{U}_{2}$, $\bar{U}_{3}$ & $\bar{u}_{2}$, $\bar{u}_{1}$ \\
$T_{(0, 1)}^{(n_1, 0, *, *, 1, 0)}$  & $\left(\rep{1},\rep{1},\rep{1},\rep{1},\rep{1}\right)$ & $-1$ & $0$ & $14$ & $21$ & $-24$ & $105$ & $45$ & $-109$ & $15$ & $14$ & $N_{31}$, $N_{37}$ & $\bar\phi_{5}$, $\bar\phi_{6}$ \\
 & $\left(\rep{1},\rep{1},\rep{1},\rep{1},\rep{1}\right)$ & $3$ & $0$ & $-4$ & $45$ & $-50$ & $181$ & $113$ & $-265$ & $35$ & $-4$ & $N_{32}$, $N_{38}$ & $\phi_{9}$, $\phi_{10}$ \\
 & $\left(\rep{1},\rep{2},\rep{1},\rep{1},\rep{1}\right)$ & $3$ & $-\tfrac{1}{2}$ & $6$ & $-17$ & $22$ & $-97$ & $-45$ & $109$ & $-15$ & $6$ & $\bar{L}_{5}$, $\bar{L}_{6}$ & $\bar{h}_{5}$, $\bar{h}_{6}$ \\
 & $\left(\rep{3},\rep{1},\rep{1},\rep{1},\rep{1}\right)$ & $3$ & $\tfrac{1}{3}$ & $6$ & $-17$ & $22$ & $-97$ & $-45$ & $109$ & $-15$ & $6$ & $D_{2}$, $D_{3}$ & $\delta_{2}$, $\delta_{3}$ \\
 & $\left(\rep{1},\rep{1},\rep{1},\rep{1},\rep{1}\right)$ & $-9$ & $0$ & $8$ & $7$ & $-8$ & $35$ & $15$ & $-39$ & $5$ & $8$ & $N_{33}$, $N_{39}$ & $\phi_{11}$, $\phi_{12}$ \\
$T_{(0, 1)}^{(n_1, 0, *, *, 1, 1)}$  & $\left(\rep{1},\rep{1},\rep{3},\rep{1},\rep{1}\right)$ & $3$ & $0$ & $-9$ & $-17$ & $23$ & $-97$ & $-49$ & $109$ & $-15$ & $-9$ & $N_{34}$, $N_{40}$ & $x_{4}$, $x_{5}$ \\
 & $\left(\rep{1},\rep{1},\rep{1},\rep{1},\rep{1}\right)$ & $3$ & $0$ & $11$ & $-17$ & $23$ & $-105$ & $-41$ & $101$ & $-15$ & $11$ & $N_{35}$, $N_{41}$ & $\bar{n}_{1}$, $\bar{n}_{2}$ \\
\hline
$T_{(1, 1)}^{(n_1, 0, 0, 0, *, *)}$  & $\left(\rep{1},\rep{1},\rep{1},\rep{1},\rep{1}\right)$ & $5$ & $0$ & $3$ & $-55$ & $65$ & $-251$ & $-143$ & $339$ & $-45$ & $4$ & $N_{42}$, $N_{51}$ & $\phi_{13}$, $\phi_{14}$ \\
 & $\left(\crep{3},\rep{1},\rep{1},\rep{1},\rep{1}\right)$ & $-3$ & $-\tfrac{1}{3}$ & $-13$ & $-7$ & $9$ & $-43$ & $-15$ & $35$ & $-5$ & $-12$ & $\bar{D}_{5}$, $\bar{D}_{6}$ & $\bar\delta_{2}$, $\bar\delta_{3}$ \\
 & $\left(\rep{1},\rep{2},\rep{1},\rep{1},\rep{1}\right)$ & $-3$ & $\tfrac{1}{2}$ & $-13$ & $-7$ & $9$ & $-43$ & $-15$ & $35$ & $-5$ & $-12$ & $L_{8}$, $L_{9}$ & $h_{5}$, $h_{6}$ \\
$T_{(1, 1)}^{(n_1, 0, 0, 1, *, *)}$  & $\left(\rep{1},\rep{1},\rep{1},\rep{1},\rep{2}\right)$ & $6$ & $0$ & $13$ & $7$ & $-8$ & $35$ & $15$ & $-43$ & $7$ & $14$ & $N_{43}$, $N_{52}$ & $y_{3}$, $y_{5}$ \\
 & $\left(\rep{1},\rep{1},\rep{1},\rep{2},\rep{1}\right)$ & $6$ & $0$ & $3$ & $7$ & $-8$ & $35$ & $15$ & $-35$ & $3$ & $4$ & $N_{44}$, $N_{53}$ & $z_{3}$, $z_{5}$ \\
 & $\left(\rep{1},\rep{1},\crep{3},\rep{1},\rep{1}\right)$ & $5$ & $0$ & $8$ & $7$ & $-8$ & $27$ & $19$ & $-35$ & $5$ & $9$ & $N_{45}$, $N_{54}$ & $\bar{x}_{4}$, $\bar{x}_{5}$ \\
 & $\left(\rep{1},\rep{1},\rep{1},\rep{1},\rep{1}\right)$ & $5$ & $0$ & $-12$ & $7$ & $-8$ & $35$ & $11$ & $-27$ & $5$ & $-11$ & $N_{46}$, $N_{55}$ & $n_{7}$, $n_{8}$ \\
$T_{(1, 1)}^{(n_1, 0, 1, 0, *, *)}$  & $\left(\rep{1},\rep{1},\rep{1},\rep{1},\rep{1}\right)$ & $5$ & $0$ & $3$ & $3$ & $-7$ & $35$ & $15$ & $-43$ & $5$ & $6$ & $N_{47}$, $N_{56}$ & $\bar\phi_{8}$, $\bar\phi_{10}$ \\
 & $\left(\rep{1},\rep{1},\rep{1},\rep{1},\rep{1}\right)$ & $-3$ & $0$ & $-17$ & $3$ & $-7$ & $35$ & $15$ & $-27$ & $5$ & $-14$ & $N_{48}$, $N_{57}$ & $\bar\phi_{9}$, $\bar\phi_{11}$ \\
$T_{(1, 1)}^{(n_1, 0, 1, 1, *, *)}$  & $\left(\rep{1},\rep{1},\rep{1},\rep{2},\rep{1}\right)$ & $6$ & $0$ & $13$ & $7$ & $-8$ & $35$ & $15$ & $-43$ & $7$ & $16$ & $N_{49}$, $N_{58}$ & $z_{4}$, $z_{6}$ \\
 & $\left(\rep{1},\rep{1},\rep{1},\rep{1},\rep{2}\right)$ & $6$ & $0$ & $3$ & $7$ & $-8$ & $35$ & $15$ & $-35$ & $3$ & $6$ & $N_{50}$, $N_{59}$ & $y_{4}$, $y_{6}$ \\
\end{longtable}
}

\bibliography{Orbifold}
\bibliographystyle{ArXiv}

\end{document}